\documentclass[11pt,reqno]{article}
\usepackage{arxiv}
\newif\ifclean
\cleanfalse 
\usepackage[letterpaper]{geometry}
\usepackage{graphicx} 
\usepackage{amsmath} %
\usepackage{amssymb} %
\usepackage{bm} %
\usepackage{physics} 
\usepackage{xcolor}
\usepackage{hyperref}
\usepackage{tikz}
\usepackage{subcaption}
\usepackage{makecell}
\usepackage{tabularx}
\usepackage{tikz}
\usepackage{comment}
\usetikzlibrary{shapes.geometric}
\usetikzlibrary{arrows.meta}
\usepackage{tikz-3dplot}

\usetikzlibrary{circuits.logic.US,circuits.logic.IEC,fit}
\newcommand\addvmargin[1]{
  \node[fit=(current bounding box),inner ysep=#1,inner xsep=0]{};
}
\usepackage{array}
\usepackage{booktabs}
\usepackage{multirow}
\usepackage{colortbl} 
\usetikzlibrary{shapes.geometric, arrows, calc, positioning}

\tikzstyle{startstop} = [rectangle, rounded corners, minimum width=3cm, minimum height=1cm,text centered, draw=black, fill=red!30]
\tikzstyle{process} = [rectangle, minimum width=3cm, minimum height=1cm, text centered, draw=black, fill=blue!30]
\tikzstyle{arrow} = [thick,->,>=stealth]
\tikzstyle{level 1} = [sibling distance=40mm]
\tikzstyle{level 2} = [sibling distance=30mm]
\title{Polyconvex Physics-Augmented Neural Network Constitutive Models in Principal Stretches}
\author{
Adrian Buganza Tepole\\
Columbia University \\
New York City, NY, USA
\And
Asghar Jadoon\\
The University of Texas at Austin \\
Austin TX, USA
\And
Manuel Rausch\\
The University of Texas at Austin \\
Austin TX, USA
\And
Jan N. Fuhg\\
The University of Texas at Austin \\
Austin TX, USA
}

\begin{document}

\date{}
\maketitle{}

\begin{abstract}
Accurate constitutive models of soft materials are crucial for understanding their mechanical behavior and ensuring reliable predictions in the design process. To this end, scientific machine learning research has produced flexible and general material model architectures that can capture the behavior of a wide range of materials, reducing the need for expert-constructed closed-form models. The focus has gradually shifted towards embedding physical constraints in the network architecture to regularize these over-parameterized models. Two popular approaches are input convex neural networks (ICNN) and neural ordinary differential equations (NODE). A related alternative has been the generalization of closed-form models, such as sparse regression from a large library. Remarkably, all prior work using ICNN or NODE uses the invariants of the Cauchy-Green tensor and none uses the principal stretches. 
In this work, we construct general polyconvex functions of the principal stretches in a physics-aware deep-learning framework and offer insights and comparisons to invariant-based formulations. 
The framework is based on recent developments to characterize polyconvex functions in terms of convex functions of the right stretch tensor $\mathbf{U}$, its cofactor $\text{cof}\mathbf{U}$, and its determinant $J$. Any convex function of a symmetric second-order tensor can be described with a convex and symmetric function of its eigenvalues. Thus, we first describe convex functions of $\mathbf{U}$ and $\text{cof}\mathbf{U}$ in terms of their respective eigenvalues using deep Holder sets composed with ICNN functions. A third ICNN takes as input $J$ and the two convex functions of $\mathbf{U}$ and $\text{cof}\mathbf{U}$, and returns the strain energy as output. The ability of the model to capture arbitrary materials is demonstrated using synthetic and experimental data.   

\end{abstract}
\section{Introduction}

\begin{figure}
\centering

\begin{tikzpicture}[
  every node/.style={rectangle, rounded corners, text width=2.8cm, align=center, draw=black, font=\small},
  level 1/.style={sibling distance=50mm, level distance=15mm},
  level 2/.style={sibling distance=35mm, level distance=10mm},
  level distance=20mm,
  edge from parent/.style={->, thick, draw}
]

\node[fill=gray!20] {Models for Soft Materials}
child { node[fill=blue!15] {Continuum level} 
  child { node[fill=green!15] {$\Psi = \Psi(\lambda_{1},\lambda_{2},\lambda_{3})$}     child { node {Ogden \cite{ogden1972large} \\
  Le Dret \cite{le1986incompressible}} }
  }
  child { node[fill=green!15] {$\Psi = \Psi(I_{1},I_{2},I_{3})$}
    child [level distance=17mm] { node {Neo-Hooke \cite{holzapfel2002nonlinear} \\
    Mooney-Rivlin \cite{holzapfel2002nonlinear} \\
    Gent \cite{gent1996new}\\
    $\vdots$}  }
  }
  }
child { node[fill=blue!15] {Micro-Mechanical}
  child [level distance=25mm] { node {3-chain \cite{james1943theory} \\
  8-chain \cite{arruda1993three} \\
    $\vdots$} 
  }
};

\end{tikzpicture}
\caption{Classification of standard constitutive models for soft materials}
\label{fig:soft_mat_model}
\end{figure}
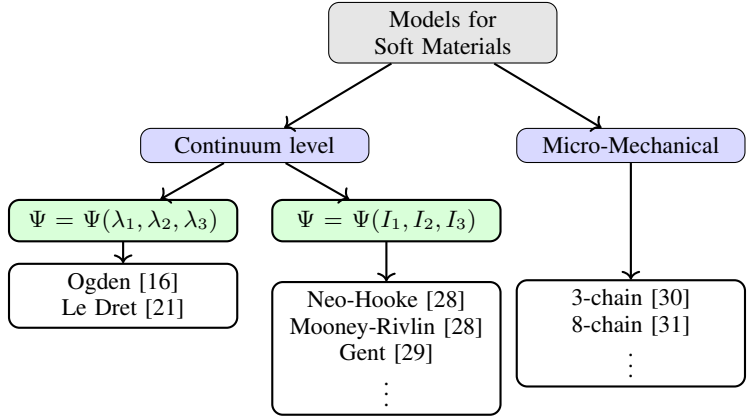

Constitutive models of soft materials are essential for selecting the appropriate material and designing structures undergoing extreme deformations. Accurate material models are also key to our understanding of the physiology and disease of soft tissue. 
Scientific machine learning research has produced flexible and general material model architectures that are able to capture a wide range of materials, reducing the need for expert-constructed closed-form models \cite{jin2023recent,fuhg2024review,tacc2024benchmarking}. 
Initial efforts in this area attempted to directly learn strain-stress data pairs using fully connected neural networks \cite{ghaboussi1998new,kirchdoerfer2016data,heider2020so,xu2021learning}. However, the focus has gradually shifted towards embedding physics constraints in the architecture such that these are satisfied by default \cite{fuhg2024review,hussain2024machine}. 
Three popular approaches are input convex neural networks (ICNN) \cite{as2022mechanics,fuhg2024extreme}, neural ordinary differential equations (NODE) \cite{tac2022data,tacc2024generative}, and generalizations of closed form models such as sparse regression from a large library or constitutive artificial neural networks \cite{linka2021constitutive,flaschel2023automated}. 
To comply with objectivity and in the case of isotropic materials, strain energy functions in terms of the isotropic invariants of the right Cauchy Green deformation tensor, $\mathbf{C}$, or its eigenvalues (the squares of the principal stretches) are used \cite{chen2022polyconvex}. Remarkably, most if not all of the data-driven approaches with ICNN or NODE are based on the invariants of $\mathbf{C}$ \cite{fuhg2024review}. Data-driven models based on sparse regression or model discovery out of a library of existing models have occasionally incorporated the Ogden model \cite{ogden1972large}, which is the only constitutive equation in terms of the squares of the principal stretches \cite{pierre2023principal}. 
The Ogden model has been highly successful in capturing the mechanical response of rubbers and biological tissues with a very simple form \cite{lohr2022introduction,destrade2022ogden}. However, surprisingly, there are barely any other examples of constitutive models in terms of principal stretches (Fig. \ref{fig:soft_mat_model}) \cite{steinmann2012hyperelastic}. An extension of the original Ogden model including a similar polynomial expansion for $\text{cof}\mathbf{C}$ was introduced in \cite{le1986incompressible,ciarlet1988three}. To our knowledge, no further developments have been attempted since \cite{steinmann2012hyperelastic}. 

In this manuscript, we were interested in building general polyconvex functions of the principal stretches in a physics-aware deep-learning framework, beyond the Ogden model. 
Our formulation relies on recent characterization of polyconvex functions in terms of convex functions of the right stretch tensor, $\mathbf{U}=\sqrt{\mathbf{C}}$, its cofactor $\text{cof}\mathbf{U}$, and its determinant $J$ \cite{steigmann2003isotropic,wiedemann2023characterization}. Any convex function of a symmetric second-order tensor can be described with a convex and symmetric function of its eigenvalues \cite{lewis1996convex}. To build convex functions of $\mathbf{U}$ and $\text{cof}\mathbf{U}$ in terms of their respective eigenvalues, we rely on a class of permutation invariant neural networks known as deep Holder sets \cite{kimura2024permutation,zaheer2017deep}. A composition of deep Holder sets with ICNNs can be used to describe general symmetric convex functions. With these building blocks, we can capture general polyconvex functions in terms of the principal stretches. 

\section{Polyconvex strain energies in terms of principal stretches}

Hyperelastic materials are described by a scalar strain energy potential, $\Psi(\bm{F})$, as a function of the deformation gradient, $\bm{F}$. This potential cannot be entirely arbitrary but rather must conform to physically meaningful constraints. The most important considerations for $\Psi$ are material symmetries, objectivity, and exhibiting some form of convexity. Ball showed that polyconvexity of $\Psi$ is sufficient to guarantee the existence of solutions to boundary value problems where the material is hyperelastic (existence of global minimizers) \cite{ball1976convexity}. While this requirement is not necessary and on occasions, it might be too restrictive \cite{schneider2017beyond}, it is convenient because it is a local condition on the strain energy as opposed to the necessary condition of quasi-convexity \cite{ball1976convexity}. Polyconvexity requires $\Psi$ to be a convex function on the extended set of arguments including $\bm{F}$, but also its cofactor and determinant,

\begin{equation}
    \Tilde{\Psi}(\bm{F}, \text{cof} \bm{F}, \det\bm{F} ) 
\label{eq:polyconvexF}
\end{equation}

In other words, $\Tilde{\Psi}$ is convex in $\mathbb{R}^{19}$ ( $\bm{F},\text{cof}\bm{F} \in \mathbb{R}^3\times \mathbb{R}^3$ and $J=\det\bm{F} \in\mathbb{R}$).  

Explicit dependence on the deformation gradient is usually avoided because $\bm{F}$ is not objective. Instead, to guarantee objectivity \textit{a priori}, $\Psi$ (or rather $\Tilde{\Psi}$) is assumed a function of the right Cauchy Green deformation tensor $\bm{C}=\bm{F}^T\bm{F}$. Additionally, consideration of material symmetries restricts the functional dependence of $\Psi(\bm{C})$. For example, for isotropic materials, the invariants $I_1=\text{tr} \bm{C}$, $I_2=\text{cof}\bm{C}$, $I_3=\text{det}\bm{C}=J^2$ are used. Polyconvexity of invariant based models  of the form $\Tilde{\Psi}(\bm{C},\text{cof}\bm{C},J)$, or more precisely $\Tilde{\Psi}(I_1,I_2,J)$ requires that $\Tilde{\Psi}$ is convex non-decreasing with respect to $I_1,I_2$ and convex with respect to $J$. Note that this is because $I_1$ is quadratic in $\bm{F}$ while $I_2$ is quadratic in $\text{cof}\bm{F}$. Therefore, composition with convex non-decreasing functions is needed to retain convexity with respect to $\bm{F}$  and  $\text{cof}\bm{F}$ \cite{boyd2004convex}.  

Alternatively, the eigenvalues of $\bm{C}$, which are the squared of the principal stretches, $\lambda_i^2$, are also invariant under rotation and therefore can be used as arguments of an isotropic strain energy potential \cite{miehe1994aspects}. However, models in terms of principal stretches are rare. The most common constitutive model in terms of principal stretches is the Ogden model \cite{steinmann2012hyperelastic,lohr2022introduction}. Only recently have there been other efforts to propose strain energies in terms of $\lambda_i^2$ \cite{dal2023data}. Yet, even recent attempts to formulate $\Psi$ in terms of principal stretches are not general enough; they only depend on isotropic functions of $\lambda_i^2$ and thus only consider convexity with respect to $\bm{F}$ and not $\text{cof}\bm{F}$ \cite{dal2023data,pierre2023principal}. Furthermore, the square of the principal stretches $\lambda_i^2$ or even powers $\lambda^{2\alpha}$ lead to convex functions of $\bm{C}$ and not necessarily with respect to $\bm{F}$. Steigmann showed a general description of polyconvex energies in terms of the stretch tensor stemming from the polar decomposition of the deformation gradient \cite{steigmann2003isotropic}, 

\begin{equation}
    \bm{F} = \bm{R} \bm{U}\quad ,
\end{equation}
with 
\begin{equation}
    \bm{U} = \sum_{i} \lambda_{i} \bm{n}_{i} \otimes \bm{n}_{i} \quad .
\end{equation}

Steigmann showed that polyconvexity of $\Psi$ can be achieved by constructing a special class of polyconvex functions of $\bm{U}$ 

\begin{equation}
    \Tilde{\Psi}(\bm{F}, \text{cof} \bm{F}, \det{(\bm{F})} ) =    \Tilde{\Psi}(\bm{U}, \text{cof} \bm{U}, \det\bm{U} ) \quad ,
    \label{eq:polyconvexU}
\end{equation}

convex on the extended domain $[\bm{U}, \text{cof} \bm{U}, \det \bm{U}]$. However, the potentials explored by Steigmann were in terms of the invariants $i_1=\text{tr}\bm{U}$, $i_2=\text{tr}(\text{cof}\bm{U})$ \cite{steigmann2003isotropic}. Namely, the potentials were of the form

\begin{equation}
    \Tilde{\Psi}(i_1, i_2, J ) 
\label{eq:polyconvex}
\end{equation}

convex on the three arguments. The result has been recently extended to more general functions of the principal stretches $\lambda_i$ by Wiedemann and Peter \cite{wiedemann2023characterization}. Considering again Eq. (\ref{eq:polyconvexU}), $\Tilde{\Psi}$ has to be convex in $[\bm{U}, \text{cof} \bm{U}, \det\bm{U}]$. Consider first a convex function $g(\bm{U})$. There exists a function $\tilde{g}(\boldsymbol{\lambda})=g(\bm{U})$ such that $\tilde{g}(\bullet)$ is convex and symmetric on its arguments, with $\boldsymbol{\lambda}\in\mathbb{R}^d$ is the vector of eigenvalues of $\bm{U}$ for dimensions $d=2,3$ \cite{rosakis1997characterization,lewis1996convex}. In our case, we consider 3-dimensional elasticity such that $\tilde{g}(\lambda_1,\lambda_2,\lambda_3)$ is a convex and symmetric function, i.e. invariant under permutations of the arguments. 

Similarly, consider the cofactor $\text{cof}\bm{U}$, we have the expansion in terms of the principal stretches

\begin{equation}
    \begin{aligned}
        \text{cof} \bm{U} &= \det{(\bm{U})} \bm{U}^{-T} = (\lambda_{i} \lambda_{2} \lambda_{3}) \left( \sum_{i} \frac{1}{\lambda_{i}} \bm{n}_{i} \otimes \bm{n}_{i} \right) \\
        &=\lambda_{2} \lambda_{3} \bm{n}_{1} \otimes \bm{n}_{1} + 
                \lambda_{1} \lambda_{3}
        \bm{n}_{2} \otimes \bm{n}_{2} + 
                \lambda_{1} \lambda_{2}
        \bm{n}_{3} \otimes \bm{n}_{3} \, .
    \end{aligned}
\end{equation}

Then, for a convex function of $\text{cof}\bm{U}$ we need a convex function $\tilde{h}(\lambda_1\lambda_2,\lambda_1\lambda_3,\lambda_2\lambda_3)$ convex and symmetric on its three arguments, the eigenvalues of $\text{cof}\bm{U}$. 

Thus, a polyconvex material can be expressed as \cite{wiedemann2023characterization}

\begin{equation}
    \Tilde{\Psi}( \lambda_1,\lambda_2,\lambda_3, \lambda_1\lambda_2,\lambda_1\lambda_3,\lambda_2\lambda_3, J)
    \label{eq:polyconvexLam}
\end{equation}

convex on its arguments, and invariant under permutations of $\lambda_1,\lambda_2,\lambda_3$ and also invariant under permutations of $\lambda_1\lambda_2,\lambda_1\lambda_3,\lambda_2\lambda_3$.

\section{Existing Hyperelastic Potentials in Terms of Principal Stretches }
As mentioned before, there are very few examples of strain energies in terms of the principal stretches. Undoubtedly the most well-known is the one proposed by Ogden \cite{ogden1972large}, which has remained one of the most accurate for modeling rubber mechanics \cite{destrade2022ogden} and has also proven useful in other domains such as modeling of soft tissue \cite{lohr2022introduction}.
The incompressible Ogden model is 
\begin{equation}
    \Psi_{\text{O}} = \sum_1^p \frac{2 \mu_p}{\alpha_p^2} (\lambda_1^{\alpha_p} + \lambda_2^{\alpha_p} + \lambda_3^{\alpha_p}-3) - p(J-1)\, .
\end{equation}
An example of a compressible Ogden model is 
\begin{equation}
    \Psi_{\text{Oc}}(\lambda_{1},\lambda_{2},\lambda_{3}) = \sum_{i=1}^{m} \frac{\gamma_{i}}{\alpha_{i}} (\overline{\lambda}_{1}^{\alpha_{i}}+\overline{\lambda}_{2}^{\alpha_{i}} +\overline{\lambda}_{3}^{\alpha_{i}}  - 3 ) + \kappa \beta^{-2} (\beta \log J + J^{-\beta} -1)
    \label{eq:Ogden_compressible}
\end{equation} 
with $\overline{\lambda}_i=J^{-1/3}\lambda_i$. This strain energy is polyconvex for $ \gamma_{i} \alpha_{i}>0$, $\abs{\alpha_{i}}>1$, $\kappa>0$, and  $\beta >0$ \cite{hartmann2003polyconvexity}.
We also consider, for completeness, an invariant-based model of a similar polynomial expansion as in Eq. (\ref{eq:Ogden_compressible}). This invariant-based model, commonly known as a generalized Ogden material, takes the form 
\begin{equation}
    \begin{aligned}
        \Psi_{\text{Oi}} = \sum_{i=1}^{m} c_{i0} (\overline{I}_{1}-3)^{i} + \sum_{j=1}^{n} c_{0j} (\overline{I}_{2}^{3/2} - 3 \sqrt{3})^{j} +  \kappa (J^{2} + J^{-2}-2)
    \end{aligned}
    \label{eq:genOgden}
\end{equation}
where $\overline{I}_{1}= \text{tr}(J^{-2/3} \bm{C})$, $\overline{I}_{2}= \text{tr}\, \text{adj}(J^{-2/3} \bm{C})$
and polyconvexity requires $c_{i0}\geq 0$, $c_{0j}\geq 0$, $\kappa\geq0$  \cite{hartmann2003polyconvexity}.
Principal stresses follow \cite{holzapfel2002nonlinear},
\begin{equation}
    \sigma_{a} = - p +\lambda_{a} \frac{\partial \Psi_{\text{dev}}}{\partial \lambda_{a}}
    \label{eq:stress_a}
\end{equation}
where $\Psi_{\text{dev}}$ is the deviatoric part of the strain energy, and $p$ can be obtained from either a volumetric strain energy $p=\partial \Psi_{\text{vol}}/\partial J$, for compressible materials, or from boundary conditions in the case of incompressible behavior $J=1$. Rubber is usually treated as incompressible, thus we present the analytical expressions for the principal stresses in common loading modes considering this assumption.
For uniaxial tension, equibiaxial tension, and pure shear the nonzero stresses are summarized in Table \ref{tab:deformationModes}.
\begin{table}
\begin{center}
\begin{tabular}{|c|| c||c |}
    \hline
Type & Deformation & Relevant experimental output\\
        \hline
                \hline 
                &   & \\
Uniaxial tension (UT)
\scalebox{0.2}{
\begin{tikzpicture}
[cube/.style={very thick,black},
			grid/.style={very thin,gray},
			axis/.style={->,blue,thick},cubeSmall/.style={thin,black}]
	\draw[cube] (0,0,1) -- (0,2,1) -- (2,2,1) -- (2,0,1) -- cycle;
	\draw[cube] (0,2,0) -- (0,2,1);
	\draw[cube] (2,0,0) -- (2,0,1);
	\draw[cube] (2,2,0) -- (2,2,1);
 	\draw[cube] (0,2,0) -- (2,2,0);
	\draw[cube] (2,2,0) -- (2,0,0);
	\draw[cubeSmall] (0.2,0.2,1) -- (0.2,1.8,1);
 	\draw[cubeSmall] (0.4,0.2,1) -- (0.4,1.8,1);
 	\draw[cubeSmall] (0.6,0.2,1) -- (0.6,1.8,1);
  	\draw[cubeSmall] (0.8,0.2,1) -- (0.8,1.8,1);
  	\draw[cubeSmall] (1.0,0.2,1) -- (1.0,1.8,1);
  	\draw[cubeSmall] (1.2,0.2,1) -- (1.2,1.8,1);
  	\draw[cubeSmall] (1.4,0.2,1) -- (1.4,1.8,1);
    \draw[cubeSmall] (1.6,0.2,1) -- (1.6,1.8,1);
    \draw[cubeSmall] (1.8,0.2,1) -- (1.8,1.8,1);
    \draw [-{Stealth[scale=2]},line width=1.5pt] (1.0,-0.2,0.5) -- (1.0,-1.5,0.5); 
    \draw [-{Stealth[scale=2]},line width=1.5pt] (1.0,2.0,0.5) -- (1.0,3.5,0.5); 
    \addvmargin{1mm};
  \end{tikzpicture}
  }
& $\bm{F} = \begin{bmatrix}
        \lambda & 0  & 0 \\
        0 & \frac{1}{\sqrt{\lambda}} & 0 \\
        0 & 0 & \frac{1}{\sqrt{\lambda}}    \end{bmatrix}$ &   $\sigma_{1} 
= -\lambda_{2} \frac{\partial \Psi}{ \partial \lambda_{2}} + \lambda_{1} \frac{\partial \Psi}{\partial \lambda_{1}}$ \\ & &\\
   \hline 
  & &  \\
Equibiaxial tension (ET) \scalebox{0.2}{
\begin{tikzpicture}
[cube/.style={very thick,black},
			grid/.style={very thin,gray},
			axis/.style={->,blue,thick},cubeSmall/.style={thin,black}]
	\draw[cube] (0,0,1) -- (0,2,1) -- (2,2,1) -- (2,0,1) -- cycle;
	\draw[cube] (0,2,0) -- (0,2,1);
	\draw[cube] (2,0,0) -- (2,0,1);
	\draw[cube] (2,2,0) -- (2,2,1);
 	\draw[cube] (0,2,0) -- (2,2,0);
	\draw[cube] (2,2,0) -- (2,0,0);
	\draw[cubeSmall] (0.2,0.2,1) -- (0.2,1.8,1);
 	\draw[cubeSmall] (0.4,0.2,1) -- (0.4,1.8,1);
 	\draw[cubeSmall] (0.6,0.2,1) -- (0.6,1.8,1);
  	\draw[cubeSmall] (0.8,0.2,1) -- (0.8,1.8,1);
  	\draw[cubeSmall] (1.0,0.2,1) -- (1.0,1.8,1);
  	\draw[cubeSmall] (1.2,0.2,1) -- (1.2,1.8,1);
  	\draw[cubeSmall] (1.4,0.2,1) -- (1.4,1.8,1);
    \draw[cubeSmall] (1.6,0.2,1) -- (1.6,1.8,1);
    \draw[cubeSmall] (1.8,0.2,1) -- (1.8,1.8,1);
    \draw [-{Stealth[scale=2]},line width=1.5pt] (1.0,-0.2,0.5) -- (1.0,-1.5,0.5); 
    \draw [-{Stealth[scale=2]},line width=1.5pt] (1.0,2.0,0.5) -- (1.0,3.5,0.5); 
    \draw [-{Stealth[scale=2]},line width=1.5pt] (2.0,1.0,0.5) -- (3.5,1.0,0.5); 
    \draw [-{Stealth[scale=2]},line width=1.5pt] (0.0,1.0,0.5) -- (-1.5,1.0,0.5); 
    \addvmargin{1mm};
  \end{tikzpicture}
  }&  
   $\bm{F} = \begin{bmatrix}
        \lambda & 0  & 0 \\
        0 & \lambda & 0 \\
        0 & 0 & \frac{1}{\lambda^2}    \end{bmatrix}$ 
  & $\sigma_{1} = \sigma_{2}= - \lambda_{3} \frac{\partial \Psi}{\partial \lambda_{3}} +\lambda_{1} \frac{\partial \Psi}{\partial \lambda_{1}}  $ \\& &   \\
   \hline 
  & &  \\
Pure shear stress (PS) \scalebox{0.2}{
\begin{tikzpicture}
[cube/.style={very thick,black},
			grid/.style={very thin,gray},
			axis/.style={->,blue,thick},cubeSmall/.style={thin,black}]
	\draw[cube] (0,0,1) -- (0,2,1) -- (2,2,1) -- (2,0,1) -- cycle;
	\draw[cube] (0,2,0) -- (0,2,1);
	\draw[cube] (2,0,0) -- (2,0,1);
	\draw[cube] (2,2,0) -- (2,2,1);
 	\draw[cube] (0,2,0) -- (2,2,0);
	\draw[cube] (2,2,0) -- (2,0,0);
	\draw[cubeSmall] (0.2,0.2,1) -- (0.2,1.8,1);
 	\draw[cubeSmall] (0.4,0.2,1) -- (0.4,1.8,1);
 	\draw[cubeSmall] (0.6,0.2,1) -- (0.6,1.8,1);
  	\draw[cubeSmall] (0.8,0.2,1) -- (0.8,1.8,1);
  	\draw[cubeSmall] (1.0,0.2,1) -- (1.0,1.8,1);
  	\draw[cubeSmall] (1.2,0.2,1) -- (1.2,1.8,1);
  	\draw[cubeSmall] (1.4,0.2,1) -- (1.4,1.8,1);
    \draw[cubeSmall] (1.6,0.2,1) -- (1.6,1.8,1);
    \draw[cubeSmall] (1.8,0.2,1) -- (1.8,1.8,1);
    \draw [-{Stealth[scale=2]},line width=1.5pt] (0.0,-0.5,0.5) -- (2.0,-0.5,0.5); 
    \draw [-{Stealth[scale=2]},line width=1.5pt] (2.0,2.5,0.5) -- (0.0,2.5,0.5); 
    \draw [-{Stealth[scale=2]},line width=1.5pt] (-.5,0.0,0.5) -- (-.5,2.0,0.5); 
    \draw [-{Stealth[scale=2]},line width=1.5pt] (2.5,2.0,0.5) -- (2.5,0.0,0.5); 
    \addvmargin{1mm};
  \end{tikzpicture}
  }& 
   $\bm{F} = \begin{bmatrix}
        \lambda & 0  & 0 \\
        0 & 1 & 0 \\
        0 & 0 & \frac{1}{\lambda}    \end{bmatrix}$ 
  & \multicolumn{1}{r|}{\makecell{ $    \sigma_{1} = - \lambda_{3} \frac{\partial \Psi}{\partial \lambda_{3}} + \lambda_{1} \frac{\partial \Psi}{\partial \lambda_{1}}$\\ \\ $    \sigma_{2} = - \lambda_{3} \frac{\partial \Psi}{\partial \lambda_{3}} + \lambda_{2} \frac{\partial \Psi}{\partial \lambda_{2}}$}} \\& &   \\
   \hline 
\end{tabular}
    \end{center}
    \caption{Studied deformation modes with relevant output}
    \label{tab:deformationModes}
\end{table}
For more general loading cases the pressure needs to be determined by imposing $J=1$ through the corresponding traction boundary conditions. 
For compressible materials, the pressure is a function of the volumetric strain energy, $p=\partial \Psi_{\text{vol}}/\partial J$. In that case, there is no constraint $J=1$ and, rather, the unknowns solved from boundary conditions are the transverse stretches. 
For uniaxial tension, we can define
\begin{equation}
\begin{aligned}
       & \lambda_{1} = \lambda, \lambda_{2}=\lambda_{3} = \lambda_t, \quad \text{with the constraint } \, \, \sigma_{22} = \sigma_{33} = 0, \\
       & \text{and we therefore obtain } \quad  \rightarrow \sigma_{22} = 0 = -p(\lambda,\lambda_t)+ \frac{1}{\lambda_{t}} \frac{\partial \Psi}{\partial \lambda_{t}}, 
\end{aligned}
\label{eq:uni_compressible}
\end{equation}
which needs to be solved for $\lambda_t$ given $\lambda$. Unfortunately, the Eq. (\ref{eq:uni_compressible}) is often nonlinear and needs to be solved iteratively, e.g., using Newton-Raphson approaches. 
For equibiaxial tension, the nonlinear equation to solve becomes with $ \lambda_{1} = \lambda_{2} =\lambda$ and $\sigma_{33} =  0$
\begin{equation}
\begin{aligned}
0 = -p(\lambda,\lambda_3)+ \frac{1}{\lambda_{3}} \frac{\partial \Psi}{\partial \lambda_{3}}, 
\end{aligned}
\label{eq:bi_compressible}
\end{equation}
where the unknown is $\lambda_3$. 
Lastly, for pure shear, i.e., $ \lambda_{1} =  \lambda, \lambda_{2}=1$  with the constraint $\sigma_{33} =  0$, the equation to solve also takes the form of Eq. (\ref{eq:bi_compressible}).
\section{Data-driven Formulation}

\begin{figure}
    \centering
    \includegraphics[width=1.0\linewidth]{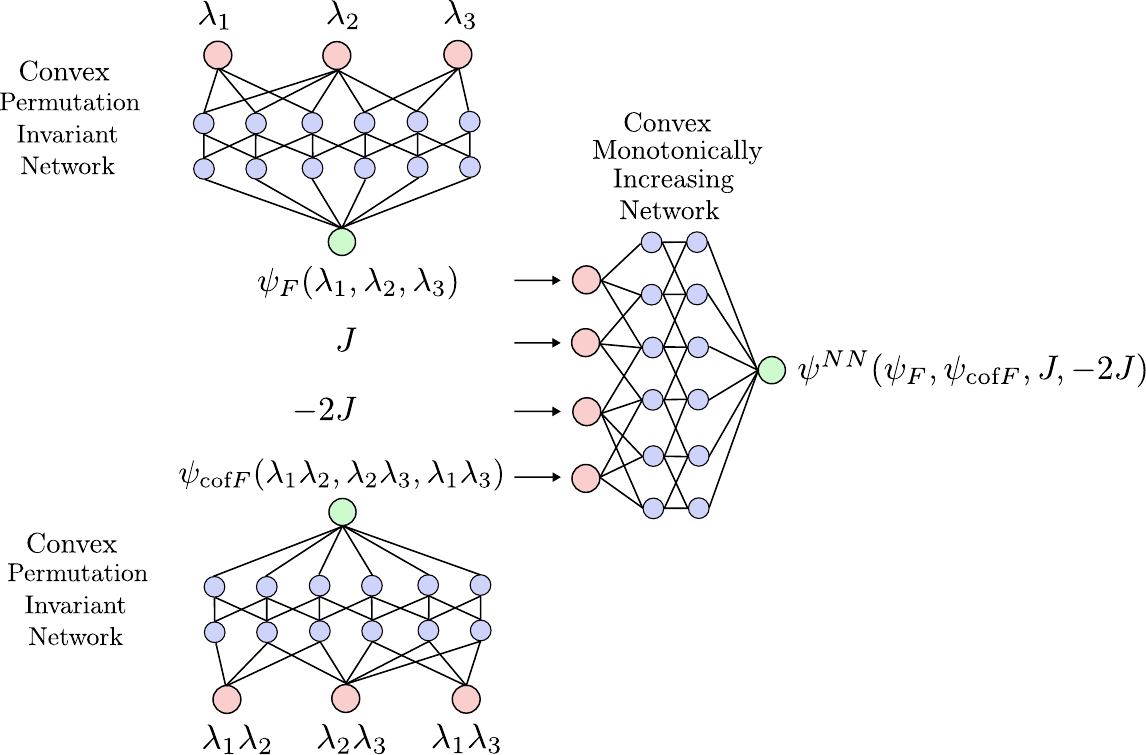}
    \caption{Schematic representation of the presented polyconvex physics-augmented neural network model with principal stretch inputs, $\lambda$-PANN.}
    \label{fig:NN_architecture}
\end{figure}

Following the general polyconvex requirement Eq. (\ref{eq:polyconvexLam}), we propose the neural network-based model illustrated in Fig. \ref{fig:NN_architecture},

\begin{equation}
\begin{aligned}
     \psi &=   \psi^{\text{NN}} (\lambda_{1},\lambda_{2},\lambda_{3},\lambda_{1}\lambda_{2},\lambda_{2}\lambda_{3},\lambda_{1}\lambda_{3}, \lambda_{1}\lambda_{2}\lambda_{3}, -2 \lambda_{1} \lambda_{2} \lambda_{3})-\psi^{\text{NN}}(1,1,1,1,1,1,1,-2)   \\
     &- \sum_{i} o_{i} (\lambda_{i} -1 ) + \psi^{\text{growth}}(J)
\end{aligned}
\end{equation}

Note that the use of $J$ and $-2J$ is needed to be able to obtain negative stress responses \cite{linden2023neural}. Since $-2J$ is convex in J, polyconvexity is preserved. We furthermore subtract the constants $o_a$ to guarantee zero stresses in the undeformed state. Also, consider the growth condition
\begin{equation}
  \psi^{\text{growth}}(J)=  \epsilon \left(\frac{1}{J} + J^{2}\right),
\end{equation}
where we choose $\epsilon=0.01$ and the general expression of stress
\begin{equation}
\begin{aligned}
        \sigma_{a} &=\frac{1}{J} \lambda_{a} \frac{\partial \psi}{\partial \lambda_{a}} 
        = \frac{\lambda_{a}}{J} \left( \frac{\partial \psi^{\text{NN}}}{\partial \lambda_{a}} - o_{a} + \frac{\partial \psi^{\text{growth}}}{\partial \lambda_{a}} \right).
\end{aligned}
\end{equation}
Given the desired normalization of the stress $\bm{\sigma}(\bm{F}=\bm{I}) = \bm{0}$ we obtain
\begin{equation}
    \begin{aligned}
        \sigma_{a}(\bm{F}=\bm{I}) &= 0 = \left( \left.\frac{\partial \psi^{\text{NN}}}{\partial \lambda_{a}}\right\vert_{\bm{F}=\bm{I}} - o_{a} + \left.\frac{\partial \psi^{\text{growth}}}{\partial \lambda_{a}}\right\vert_{\bm{F}=\bm{I}} \right) \\
        \rightarrow o_{a} &= \left.\frac{\partial \psi^{\text{NN}}}{\partial \lambda_{a}}\right\vert_{\bm{F}=\bm{I}}  + \left.\frac{\partial \psi^{\text{growth}}}{\partial \lambda_{a}}\right\vert_{\bm{F}=\bm{I}}. 
    \end{aligned}
\end{equation}

To satisfy the convexity of $\psi^{\text{NN}}$ and also the desired permutation invariances, we split $\psi^{\text{NN}}$ into three neural networks as seen in Fig. \ref{fig:NN_architecture}. Two initial neural networks are denoted $\Psi_F$ and $\Psi_{\text{cof}F}$ and produce, as the notation suggests, convex functions of $\bm{U}$ and $\text{cof}\bm{U}$ by having a permutation invariant function of the eigenvalues $\lambda_1,\lambda_2,\lambda_3$  and 
$\lambda_1\lambda_2,\lambda_1\lambda_3,\lambda_2\lambda_3$, respectively. 

For the permutation invariant neural network we adopt the framework by Kimura et al., which is an extension of the Deep Set framework by Zaheer et al \cite{kimura2024permutation,zaheer2017deep}. The Hölder’s Power Deep Sets take the form 

\begin{equation}
    g(x_1,x_2,x_3) = \rho \left( \left[\sum_i \phi (x_i)^p \right]^{1/p}\right)
    \label{eq:power_deep_set},
\end{equation}

where $\phi$ is an input convex neural network, which thus represents a convex function of $x_i$. Note that the $p-norm$ type approach gives general permutation invariant functions depending on $p$. For $p=1$ we get just the mean $(\phi(x1)+\phi(x_2)+\phi(x_3))/3$ while for $p=\infty$ you get the maximum function $\max(\phi(x_1),\phi(\lambda_2),\phi(\lambda_3))$, which are two common symmetric functions. The function $\rho$ is a convex monotonically increasing function of one scalar argument. The monotonicity constraint is needed to maintain convexity with respect to $x_i$ \cite{boyd2004convex}. 

After evaluating $\Psi_F, \Psi_{cofF}$, we pass these outputs through an ICNN to obtain convex non-decreasing functions of $\mathbf{U}$ and $\text{cof}\mathbf{U}$. In addition to taking $\Psi_F and \Psi_{\text{cof}F}$ as inputs, this last ICNN also takes in $J$ and $-2J$ to capture arbitrary convex functions of $J$. Thus, the $\lambda$-PANN architecture in Fig. \ref{fig:NN_architecture} describes a broad space of polyconvex functions in terms of principal stretches.

\section{Results}

We first test whether the data-driven formulation is able to capture synthetic Ogden models based on the strain energy Eq. (\ref{eq:Ogden_compressible}). We randomly sampled parameters for Eq. (\ref{eq:Ogden_compressible}) and reported them in Table \ref{tb:initialGuessSens}. The parameters are sampled from the following  uniform distributions $\beta \sim \mathcal{U}(1,2)$, $\mu  \sim  \mathcal{U}(-5,5) $ and $\alpha \sim \mathcal{U} \left( [-5,-1]\bigcup [1,5] \right)$ and the sign of the $\mu_i,\alpha_i$ pairs matched to ensure the polyconvex conditions  $ \mu_{i} \alpha_{i}>0$, $\abs{\alpha_{i}}>1$, $\kappa>0$, and  $\beta >0$ .  

\begin{table}[h!]
\centering
\caption{Ogden parameters}
\resizebox{0.7\textwidth}{!}{%
\begin{tabular}{ c  c c c c c c c}
\toprule
\rowcolor{gray!10}
$\mu_{1}$ &$\mu_{2}$ & $\mu_{3}$&$\alpha_{1}$&$\alpha_{2}$&$\alpha_{3}$ & $\kappa$& $\beta$\\
\midrule
-2.933&0.101&2.832 & -1.019&3.711&2.08&  47.592&1.963 \\
0.621&-1.396&0.775 & 4.878&-3.244&1.075&  53.929&1.241
\\
1.786&-2.949&1.163 & 1.263&-1.174&2.581&  37.179&1.206 \\ 
3.62&-2.375&-1.244 & 1.268&-1.29&-1.796&  17.555&1.109 \\
-0.866&0.375&0.491 & -1.672&3.934&2.634&  13.146&1.938 \\
1.827&-2.39&0.563 & 2.061&-1.537&2.208&  15.135&1.325\\
0.294&0.963&-1.258 & 2.62&1.593&-3.729&  12.592&1.652 \\
-1.544&2.315&-0.771 & -2.943&1.548&-2.374&  22.143&1.307 \\
-0.195&-2.232&2.427 & -3.508&-1.861&1.604&  27.666&1.109 \\
-0.052&-2.509&2.561 & -1.535&-1.568&1.874&  22.642&1.256\\
\bottomrule
\end{tabular}%
}
\label{tb:initialGuessSens}
\end{table}

To generate the data we then randomly sample deformation gradients $\mathbf{F}$. To fill a broad deformation space we do rejection sampling in the space of deformation gradients, 
\begin{equation}
    F_{ij} \in \begin{cases}
        1 + \mathcal{U}(-\delta, \delta), &\text{if} \quad i=j \\
        \mathcal{U}(-\delta, \delta), & \text{else},
    \end{cases}
\end{equation}
where a sample is accepted if 
\begin{equation}\label{eq::ConstrainedSpaceCondition}
\begin{aligned}
          - 4 i_{1}^{3} i_{3} + i_{1}^{2} i_{2}^{2}  + 18 i_{1} i_{2} i_{3} - 4 i_{2}^{3} - 27 i_{3}^{2}&&\geq 0, 
\end{aligned}
\end{equation}
with
\begin{equation}
    \begin{aligned}
        i_{1} &=  \lambda_{1} + \lambda_{2} + \lambda_{3}, \quad 
        i_{2} = \lambda_{1} \lambda_{2} +\lambda_{1} \lambda_{3} + \lambda_{2} \lambda_{3} \, \quad 
        i_{3}=\lambda_{1} \lambda_{2} \lambda_{3}.
    \end{aligned}
    \label{eq:sampleF}
\end{equation}
The latter is to ensure that the deformation is physical, i.e., $J>0$ and all principal stretches positive \cite{fuhg2022physics}.
We generate $200$ points for training with $\delta=0.2$ and $500$ points for testing with $\delta=0.3$.
After sampling the deformation gradient, we evaluate the strain energy and stresses according to Eqs. (\ref{eq:Ogden_compressible}). This allows us to build datasets for training and testing
\begin{equation}
    \mathcal{D} = \lbrace \bm{F}_{i}, \bm{\sigma}_{i} \rbrace_{i=1}^{N}.
\end{equation}


Each of the three networks that make up the $\lambda$-PANN architecture uses two layers with 10 neurons and a Softplus activation function. This results in 1285 parameters. We utilize the Adam optimizer \cite{kingma2014adam} with a learning rate of $10^{-3}$.
The median training performances of the proposed data-driven architectures ($\lambda$-PANN) across all 10 sampled Ogden models Eq. (\ref{eq:Ogden_compressible}), with parameters in Table \ref{tb:initialGuessSens}, are depicted in Fig. \ref{fig:case1_ortho_loss}. We tested the effect of the power $p$ in the $p$-norm-like expression at the core of the permutation invariant neural network, the  Hölder’s power deep sets Eq. (\ref{eq:power_deep_set}). Over 100k epochs, the training loss sharply drops and stabilizes at a minimum,  indicating good convergence to an accurate fit with losses $\sim 10^{-4}$. The best performance is achieved with the higher exponent $p=3$. For comparison, Fig. \ref{fig:case1_ortho_loss} also shows the training loss of our (and others') previously developed invariant-based polyconvex neural network ($\mathcal{I}$-PANN) \cite{linden2023neural,jadoon2024inverse}. To be consistent with the $\lambda$-PANN architecture we choose two layers with 23 neurons in the $\mathcal{I}$-PANN with a Softplus activation function. This results in 1243 parameters and therefore around the same as used for the $\lambda$-PANN. We use the same learning rate as before.

\begin{figure}
    \begin{subfigure}{0.5\linewidth}
    \centering
        \includegraphics[scale=0.24]{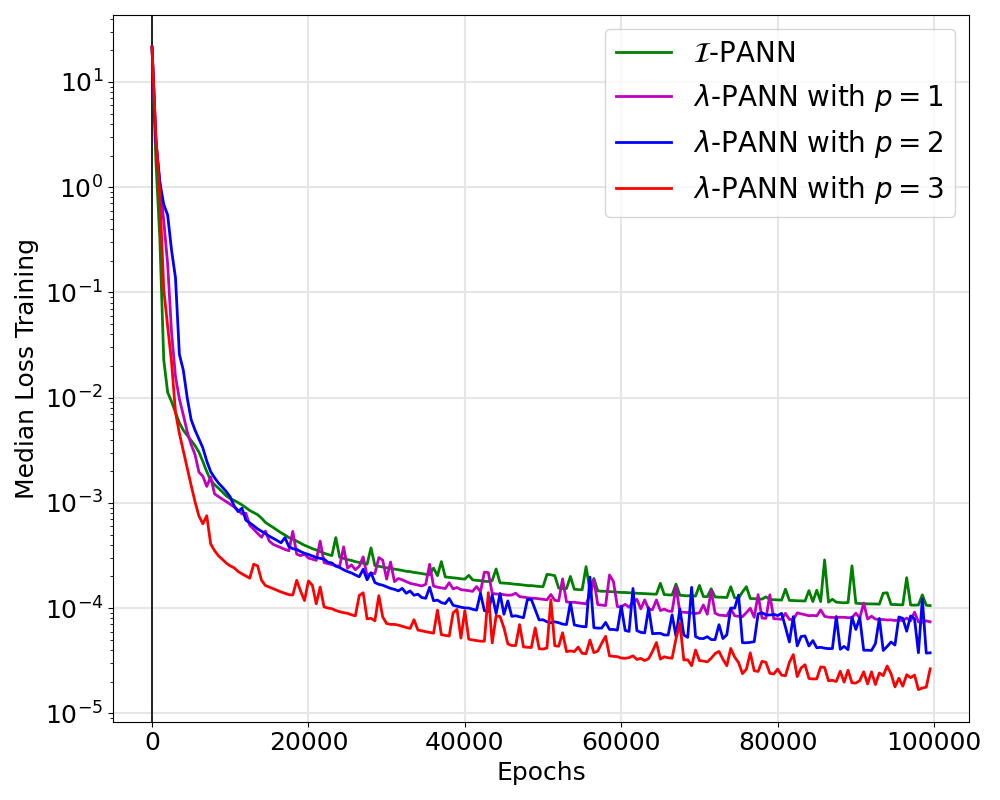}
    \caption{Training loss $\delta=0.2$}\label{fig:case1_ortho_loss}
    \end{subfigure}
        \begin{subfigure}{0.5\linewidth}
            \centering
        \includegraphics[scale=0.24]{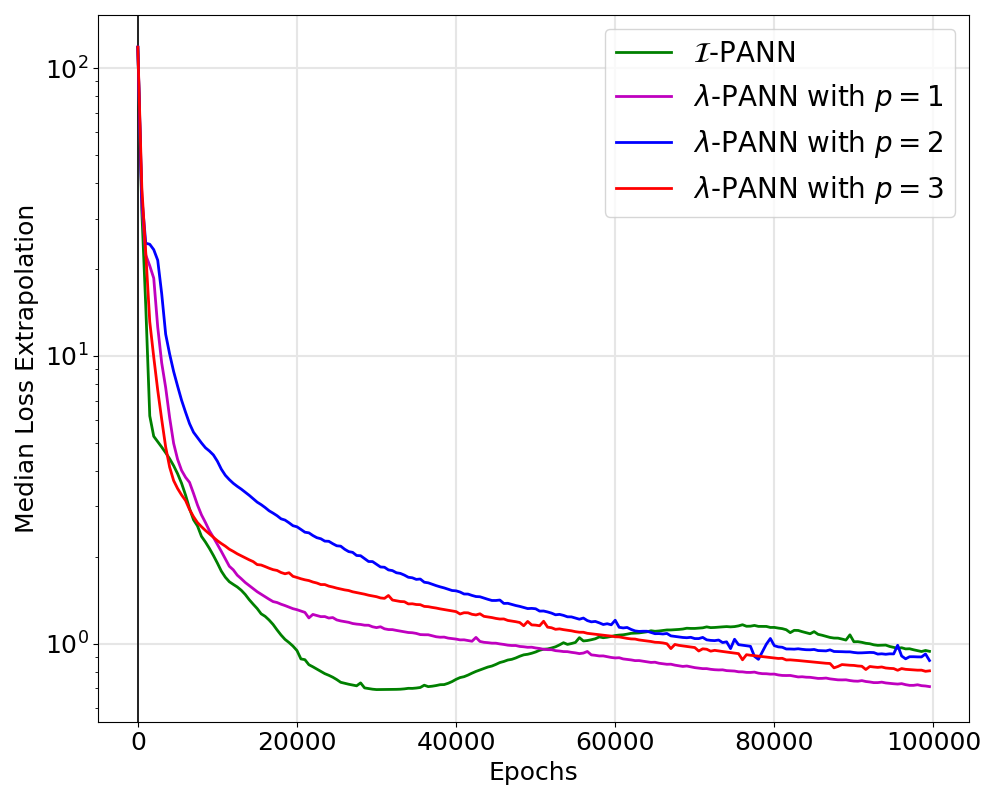}
    \caption{Extrapolation loss $\delta=0.3$}\label{fig:case1_ortho_stress}
    \end{subfigure}
    \caption{Ogden model median losses over $10$ random models as specified in Table \ref{tb:initialGuessSens}. a) Training loss; b) Extrapolation loss. }
    \label{ortho_case1}
\end{figure}

While the $\mathcal{I}$-PANN performs well in the training set, the losses are higher compared to the $\lambda$-PANN model. Interestingly, tested on an extrapolation dataset (deformations beyond the training region), the architecture based on principal stretches, $\lambda$-PANN, performs better than the invariant-based architecture, $\mathcal{I}$-PANN. Specifically, the invariant approach, $\mathcal{I}$-PANN, shows signs of over-fitting: the extrapolation loss first decays but then rises as training progresses (Fig. \ref{ortho_case1}b).

\begin{figure}
    \begin{subfigure}{0.5\linewidth}
    \centering
        \includegraphics[scale=0.24]{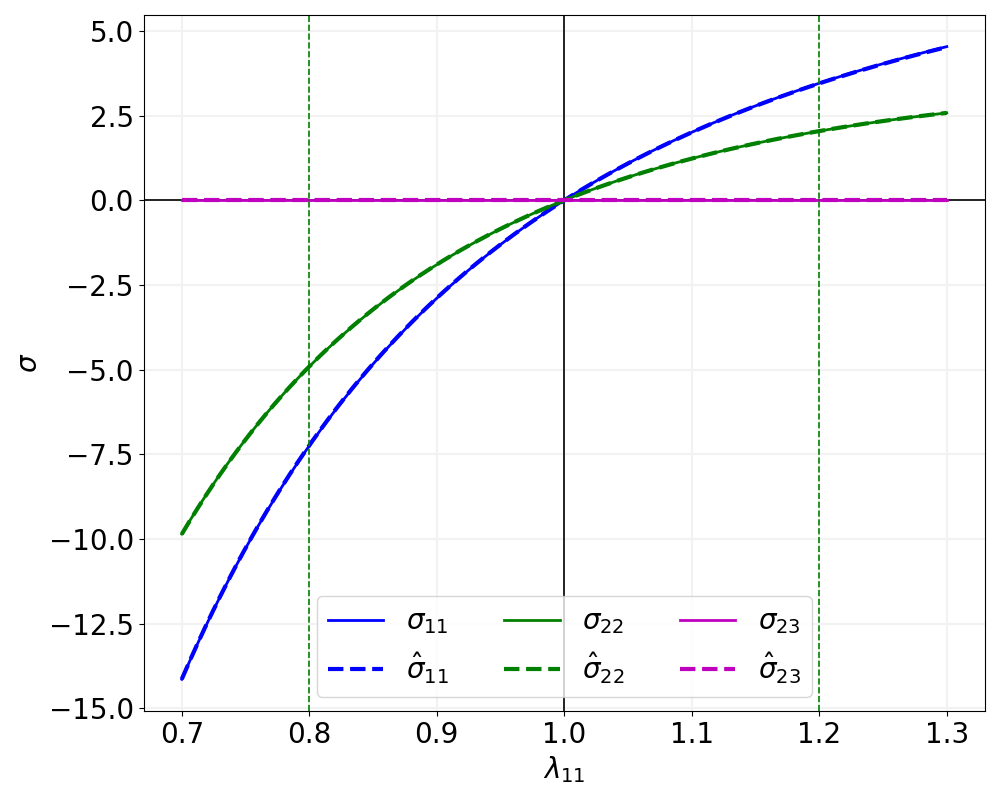}
    \caption{$p=1$}\label{fig:case1_ortho_stress-strain}
    \end{subfigure}
        \begin{subfigure}{0.5\linewidth}
        \centering
        \includegraphics[scale=0.24]{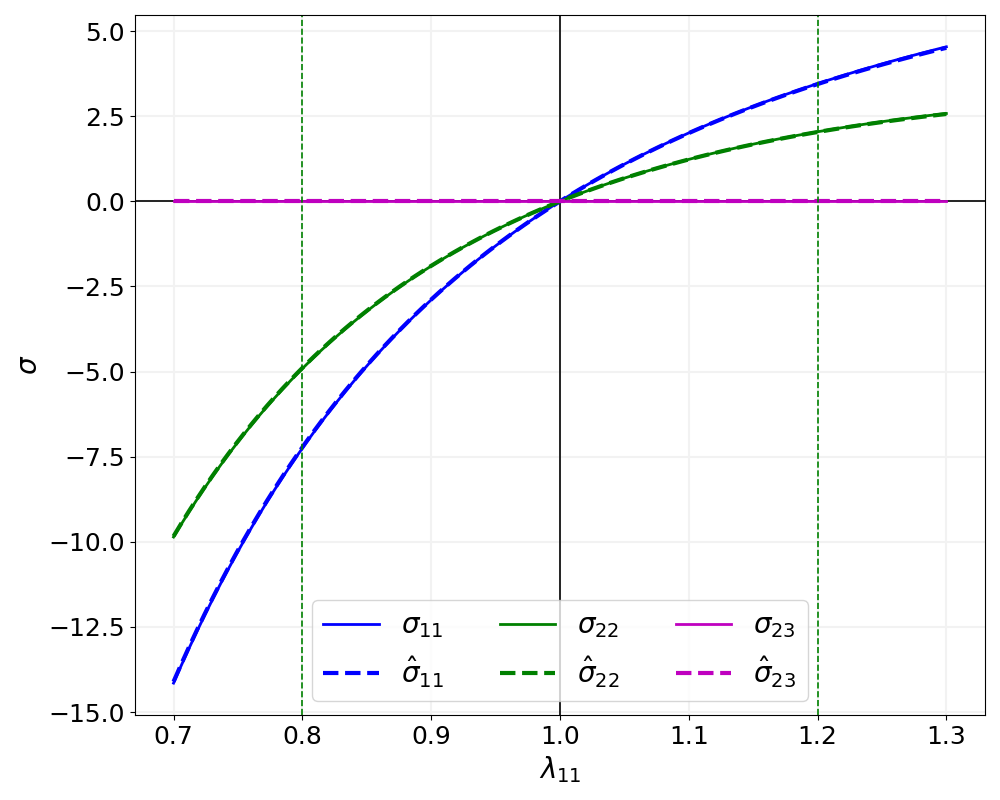}
    \caption{$p=3$}\label{fig:case1_ortho_stress}
    \end{subfigure}
    \caption{Visualizing response of $\lambda$-PANNs on Dataset 5 of $\lambda$-Ogden set. Dotted green lines indicate the training domain of $20\%$. Dashed lines denote model predictions and solid lines denote ground truth response. }
    \label{ortho_case1_stress-strain}
\end{figure}

Fig. \ref{ortho_case1_stress-strain} shows the evaluation of the trained $\lambda$-PANN model corresponding to the Ogden material Dataset 5 in Table \ref{tb:initialGuessSens}. Results for the two hyper-parameter values, $p=1$ and $p=3$, are shown, together with the ground truth evaluation. Even though from Fig. \ref{ortho_case1}, the higher exponent in the permutation invariant neural network, $p=3$, leads to better than the case $p=1$, visualizing the results in terms of stress-strain responses under deformations within and outside the training region in Fig. \ref{fig:case1_ortho_stress}, both models accurately describe the true Ogden model from Dataset 5 with no apparent difference between the two.

We next investigated if the principal stretch data-driven model would perform equally well in a synthetic dataset generated with an invariant model. 
We used the generalized Ogden model in Eq. (\ref{eq:genOgden}) and generated stress-strain data by first randomly sampling material parameters reported in Table \ref{tb:paramGeneralOgden}. 
We uniformly sample the number of active terms to be either two or three, i.e., $m  \sim \mathcal{U}_{D}(2,3)$ and $n  \sim \mathcal{U}_{D}(2,3)$ where $\mathcal{U}_{D}$ is the discrete uniform distributions. We then sample the coefficient values with $c_{i0} \sim \mathcal{U}(0.1,3.0)$ and $c_{0j} \sim \mathcal{U}(0.1,3.0)$ as well as $\kappa \sim \mathcal{U}(0.1,1.0)$. 
We used the same deformation gradients for training and testing sampled with Eq. (\ref{eq:sampleF}) for the $\lambda$-Ogden dataset. All other (hyper)parameters, such as the number of parameters of the two networks and the learning rate, were also left unchanged.

\begin{table}[h!]
\centering
\caption{Generalized Ogden parameters}
\resizebox{0.6\textwidth}{!}{%
\begin{tabular}{ c  c c c c c c}
\toprule
\rowcolor{gray!10}
$c_{10}$ &$c_{20}$ & $c_{30}$&$c_{01}$&$c_{02}$&$c_{03}$& $K$\\
\midrule
1.583 & 0.133 & - & 2.9&  0.342 & 0.248 & 0.873 \\
1.302 & 0.261 & 0.246 & 0.668 & 0.245 & 0.143 &  0.831 \\ 
0.875 & 0.181 & - & 1.433 & 0.312 & 0.229 &  0.804 \\ 
0.786 &  0.577&- & 1.268&1.334 & -&  0.86 \\
1.221 & 0.126&- & 2.874 & 0.228&-&  0.493 \\
0.909 & 0.318 & 0.18 & 2.604 & 0.238 & - &  0.743\\
2.892 & 0.248&- & 0.869 & 0.312 &  0.246&  0.931\\
0.567 & 0.533 & 0.408 &0.253 & 0.236&-&  0.954\\
0.967 & 0.906 & 0.241 & 0.341 & 0.185 &-&0.968\\
2.234 & 0.13&- & 2.762 & 0.109 &-&  0.391\\
\bottomrule
\end{tabular}%
}
\label{tb:paramGeneralOgden}
\end{table}

\begin{figure}
    \begin{subfigure}{0.49\linewidth}
    \centering
        \includegraphics[scale=0.24]{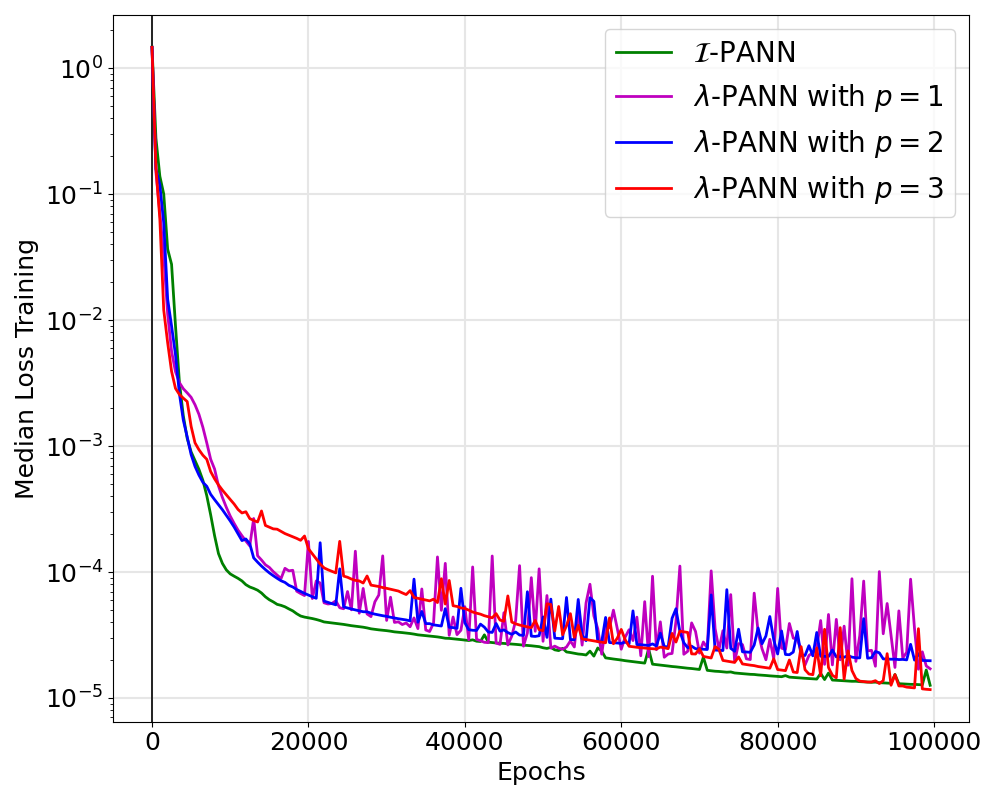}
    \caption{Training loss}\label{fig:case2_ortho_loss}
    \end{subfigure}
        \begin{subfigure}{0.49\linewidth}
        \centering
        \includegraphics[scale=0.24]{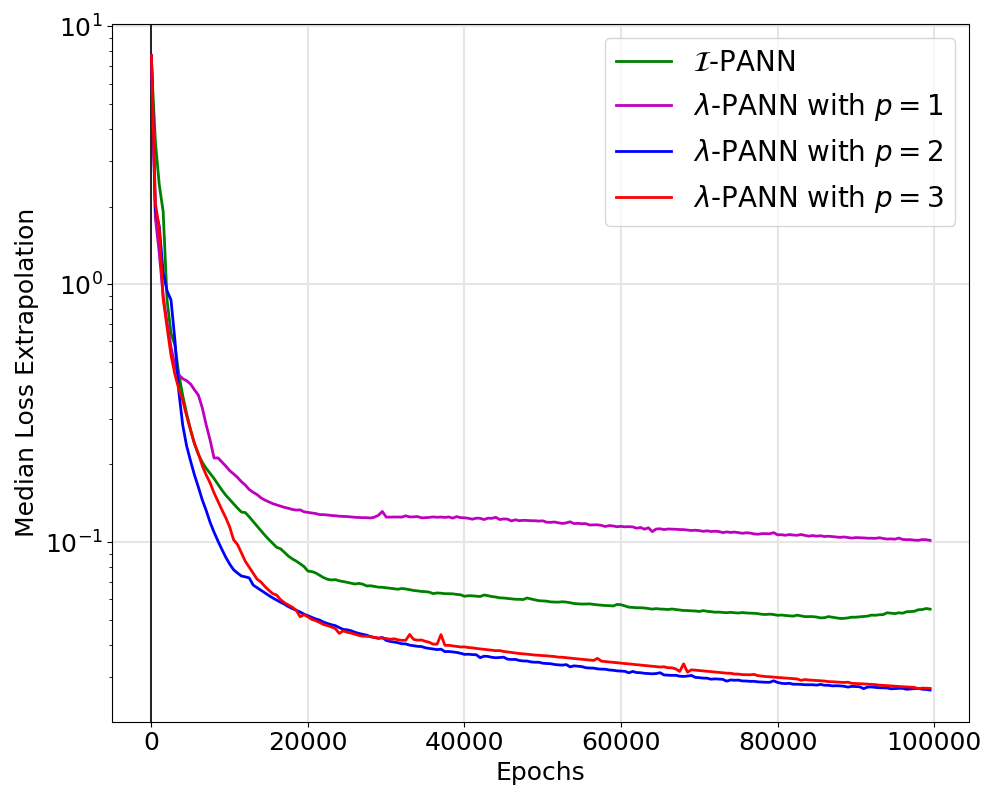}
    \caption{Extrapolation loss}\label{fig:case1_ortho_stress}
    \end{subfigure}
    \caption{Generalized Ogden model  median losses over $10$ random models as specified in Table \ref{tb:paramGeneralOgden}. a) Training loss; b) Extrapolation loss. }
    \label{ortho_case2}
\end{figure}

We trained again the same $\lambda$-PANNs as before, with this new dataset. That is, we kept the same architecture (two layers with $10$ neurons for each of the networks  ), and the three p-norm exponents ($p=1,2,3$), and trained for 100k epochs (Fig. \ref{fig:case2_ortho_loss}). For comparison, we also trained the invariant-based data-driven model $\mathcal{I}$-PANN \cite{klein2022polyconvex,linden2023neural,fuhg2022learning}. Because the ground truth data comes from an invariant base, the $\mathcal{I}$-PANN model outperforms the $\lambda$-PANNs during training. Nevertheless, both models achieve excellent losses on the order of $10^{-5}$. Interestingly, in the extrapolation dataset, with deformations in which the maximum principal stretch exceeds 1.2,  the   $\lambda$-PANNs with $p=2,3$ show notably lower losses compared to the $\mathcal{I}$-PANN model. None of the models show signs of over-fitting.

Having established the ability of the proposed $\lambda$-PANN architecture to capture stress-strain data synthetically generated with the Ogden model or the generalized invariant model, we further evaluated the effect of hyperparameter tuning and ablation of the various architecture components. The exponent in the $p$-norm evaluation at the center of the permutation-invariant neural network was already evaluated in Figs. \ref{ortho_case1} and \ref{ortho_case2}, which showed that $p=3$ had the best performance in both training and validation datasets. 

\begin{figure}
    \begin{subfigure}{0.5\linewidth}
    \centering
        \includegraphics[scale=0.24]{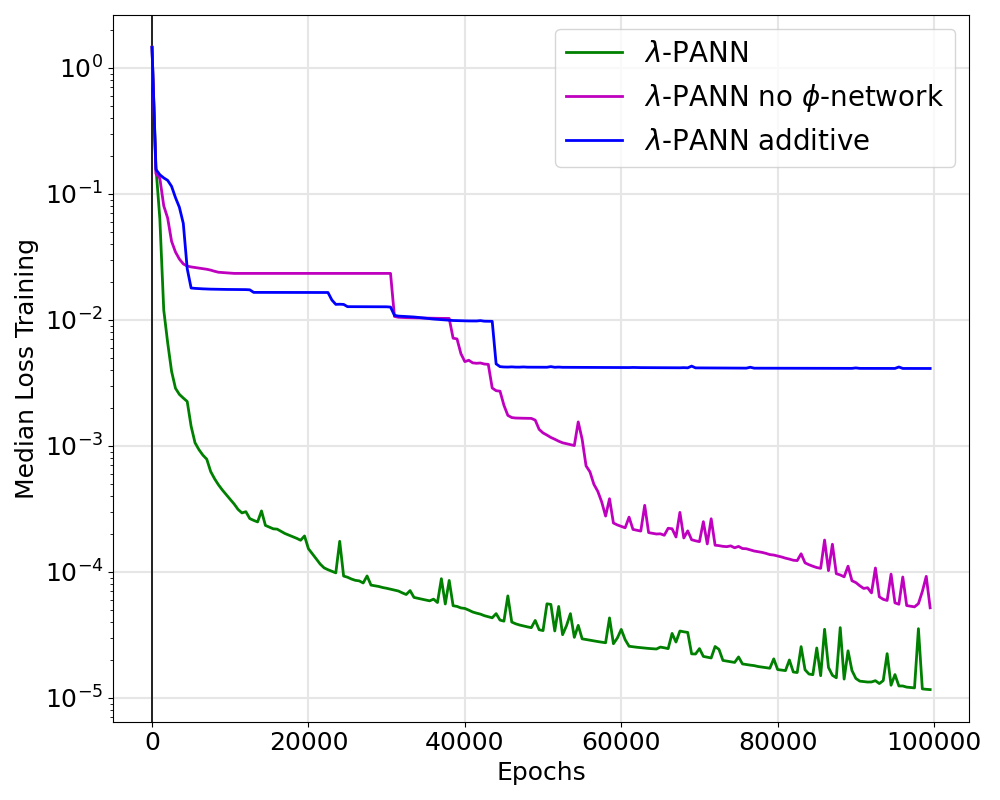}
    \caption{$p=3$}\label{fig:case1_ortho_loss}
    \end{subfigure}
        \begin{subfigure}{0.5\linewidth}
        \centering
        \includegraphics[scale=0.24]{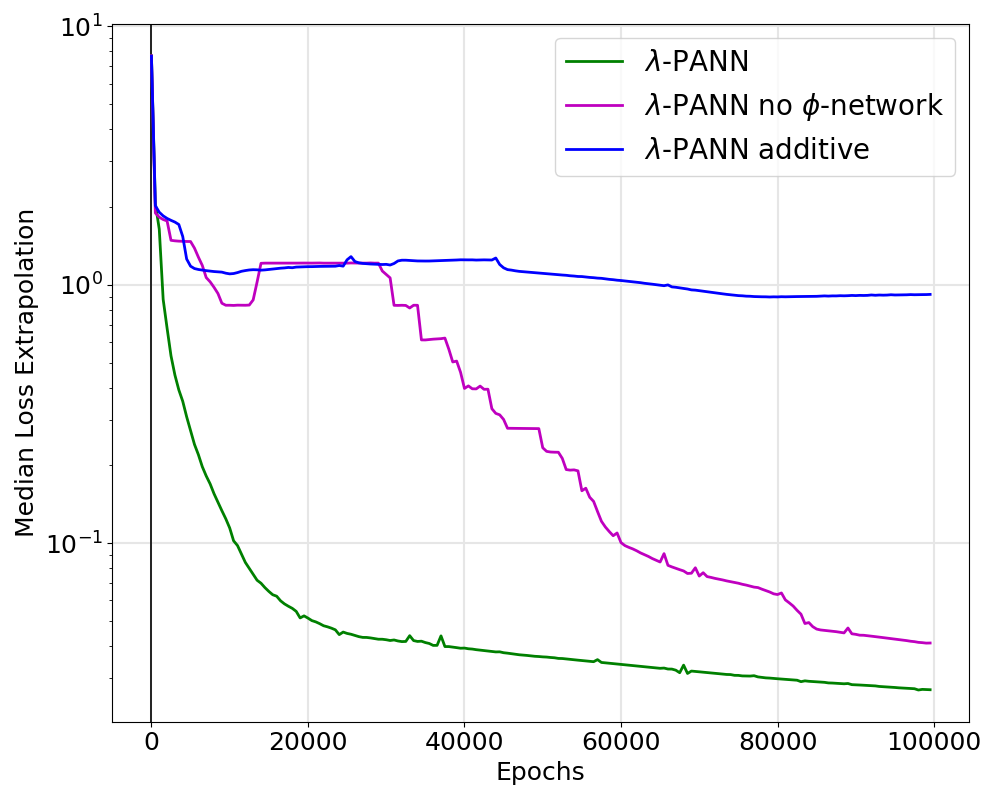}
    \caption{$p=3$}\label{fig:case1_ortho_stress}
    \end{subfigure}
    \caption{Influence of network architecture for $p=3$ for the Generalized Ogden data}
    \label{fig:ablation}
\end{figure}

Fig. \ref{fig:ablation} shows the effect of nesting multiple permutation invariant and input-convex neural networks. The more general framework illustrated in Fig. \ref{fig:NN_architecture} consists of two permutation invariant neural networks, one for capturing convex functions of $\mathbf{U}$ and one for capturing convex functions of $\text{cof}\mathbf{U}$. These permutation invariant neural networks are based on Hölder’s power deep sets Eq. (\ref{eq:power_deep_set}) in which an ICNN $\phi$ is nested within another ICNN $\rho$ which is also non-decreasing. Thus, one of the ablation studies we performed was to disregard the $\phi$ network but otherwise keep the rest of the architecture. The $\lambda$-PANN with no $\phi$ network has a much slower convergence than the original $\lambda$-PANN, as seen in Fig. \ref{fig:ablation}a. Nevertheless, after 100k epochs, the loss in the $\lambda$-PANN with no $\phi$ network approaches values seen in the original $\lambda$-PANN. Similar trends can be seen in the validation loss, Fig. \ref{fig:ablation}b. Thus, the $\phi$ network in Eq. (\ref{eq:power_deep_set}) provides flexibility to the architecture that helps it attain lower training and validation losses. 

A second ablation study explored the relevance of the last ICNN, $\psi^{NN}(\psi_F,\psi_{\text{cof}F},J,-2J)$, which takes in as inputs the outputs of the permutation invariant networks $\psi_F$ and $\psi_{\text{cof}F}$. We tested whether such nonlinear nesting would be required for performance. Thus, as an alternative, we tested an additive framework of the form $\psi_{F}+ \psi_{\text{cof}F}+\psi_J(J)$ for which $\psi_J$ is an ICNN which is just a function of $J$. The performance of this additive strain energy function is shown in Fig. \ref{fig:ablation}. Both training and validation losses of the additive neural network are significantly worse than the proposed $\lambda$-PANN which includes the nonlinear nesting $\psi^{NN}(\psi_F,\psi_{\text{cof}F},J,-2J)$.

\begin{figure}
    \centering
    \includegraphics[width=0.5\linewidth]{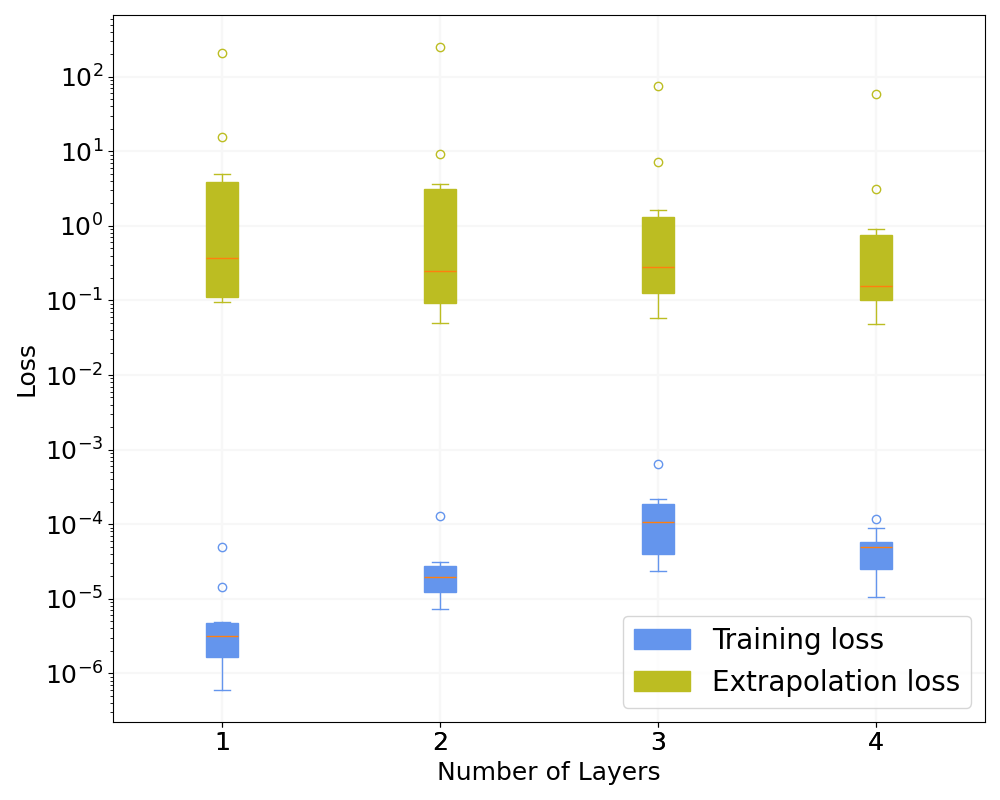}
    \caption{Impact of increasing the network size of $\lambda$-PANNs for $p=3$ for the Ogden dataset described in Table \ref{tb:initialGuessSens}. Each of the three subnetworks has 20 neurons per layer where the number of layers sees an increase from 1 to 4. }
    \label{fig:hyperparam_tuning}
\end{figure}

The last hyperparameter tuning we investigated was to vary the number of parameters by changing the number of layers to $\{1,2,3,4\}$. As seen in Fig. \ref{fig:hyperparam_tuning}, when each of the neural networks of the $\lambda$-PANN has only one layer, the training loss is the minimum, but the extrapolation loss is at its maximum. Four layers seems to be optimal for the cases considered. However, all four cases had satisfactory training losses $\leq10^{-4}$ and extrapolation losses $\leq 10^0$.

We used the optimal $\lambda$-PANN architecture identified with the synthetic datasets and applied it to the experimental rubber datasets from Treloar \cite{treloar1944stress} and Heuillet \cite{heuillet1997modelisation}. To regularize our networks and prevent overfitting due to the limited number of experimental data points, we have used L$^{0}$-regularization, see \cite{louizos2017learning,fuhg2024extreme}. We have used the same hyperparameters for this regularization as in our previous work \cite{FUHG2024105837} and a regularization factor of $10^{-4}$ between the data loss and the L$^{0}$-loss term.

\begin{figure}
    \begin{subfigure}{0.5\linewidth}
    \centering
        \includegraphics[scale=0.24]{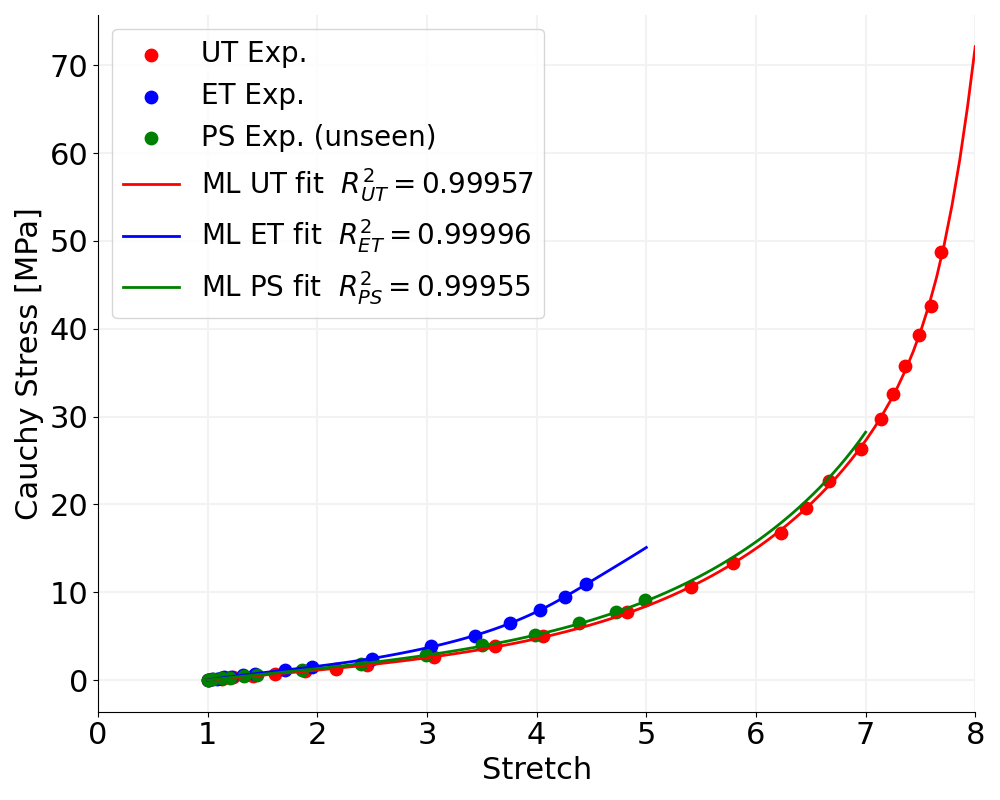}
    \caption{Treloar dataset response }\label{fig:case1_ortho_loss}
    \end{subfigure}
        \begin{subfigure}{0.5\linewidth}
        \centering
        \includegraphics[scale=0.24]{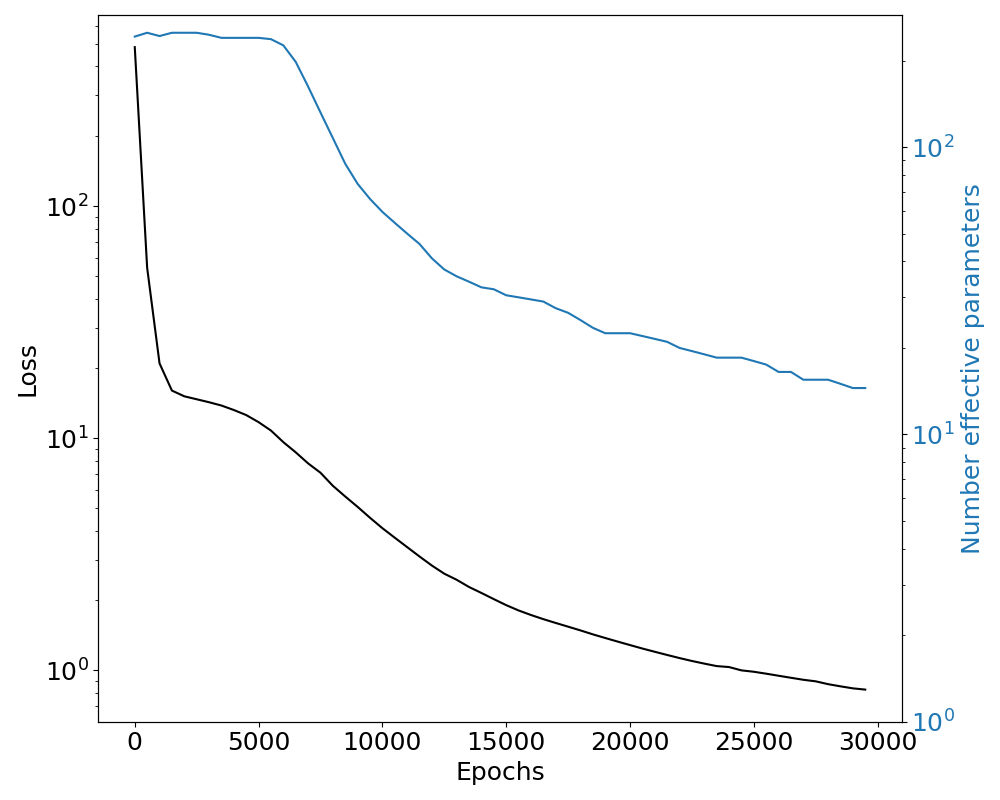}
    \caption{Treloar dataset training evolution }\label{fig:case1_ortho_stress}
    \end{subfigure}
        \begin{subfigure}{0.5\linewidth}
        \centering
        \includegraphics[scale=0.24]{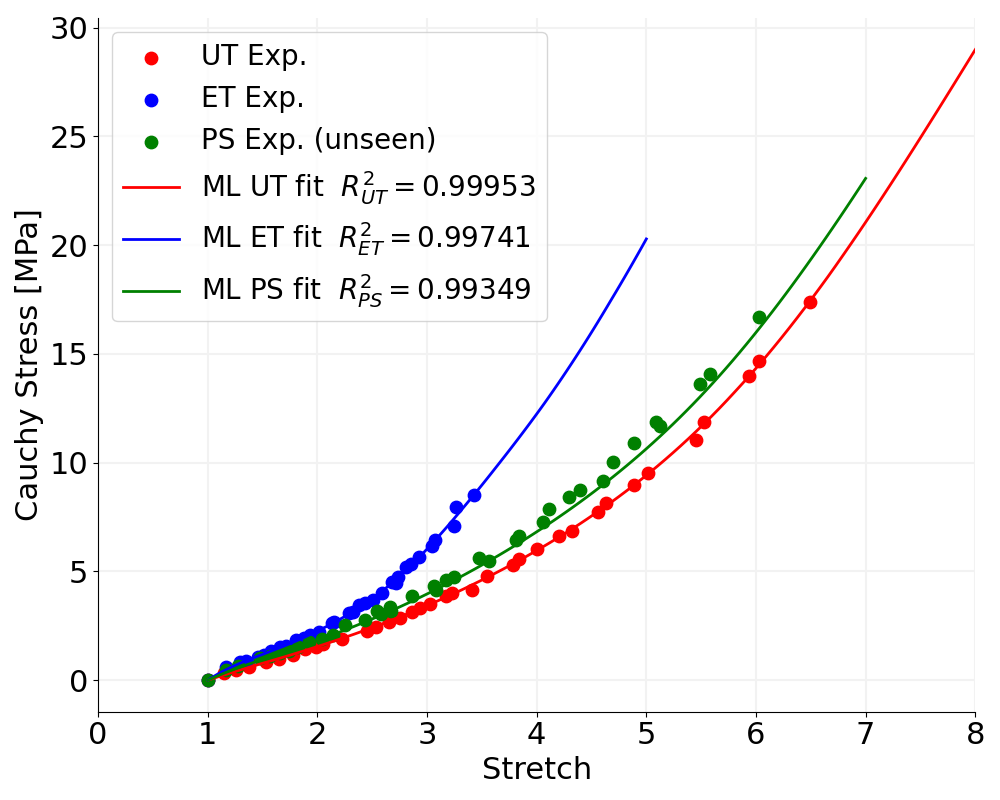}
    \caption{Heuillet dataset response }\label{fig:case1_ortho_invDir1}
    \end{subfigure}
        \begin{subfigure}{0.5\linewidth}
        \centering
        \includegraphics[scale=0.24]{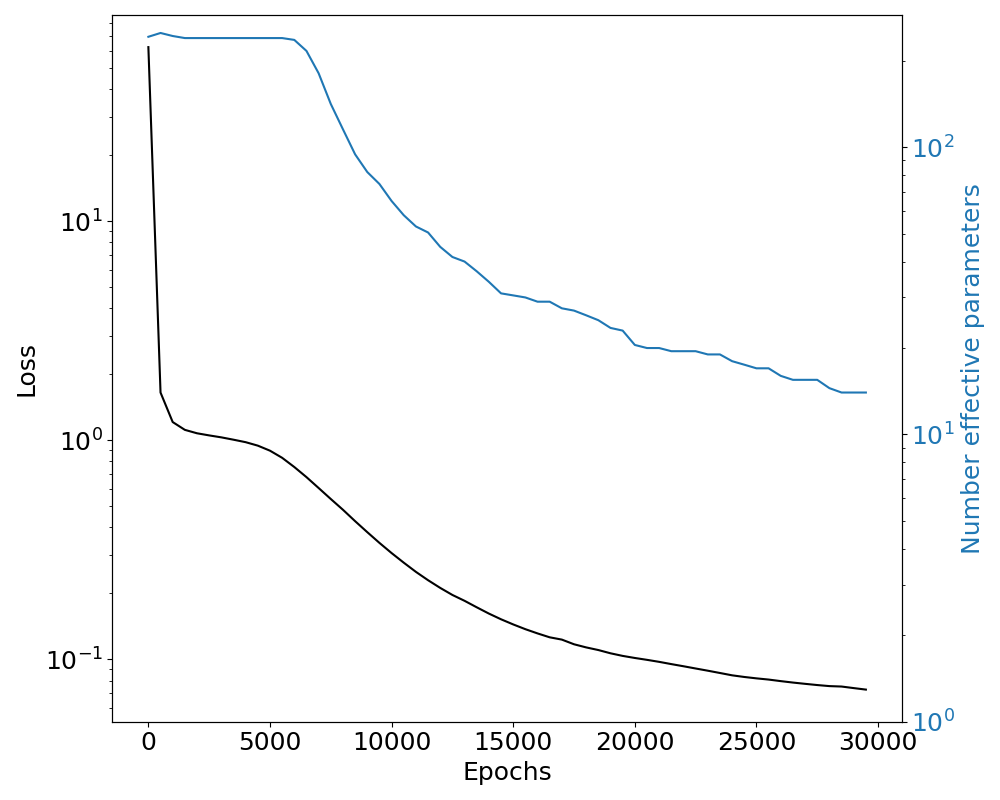}
    \caption{Heillet dataset training evolution}\label{fig:case1_ortho_invDir2}
    \end{subfigure}
    \caption{Applying $\lambda$-PANNs ($p=3$) on experimental rubber data. (a,b) Treloar data set \cite{treloar1944stress} stress response and loss and parameter evolution over training iterations, (c,d) Heuillet data set \cite{heuillet1997modelisation} stress response and loss and parameter evolution over training iterations}
    \label{fig:rubber_experiment}
\end{figure}

These datasets contain stress-strain curves under uniaxial tension (UT), pure shear (PS), and equibiaxial tension (ET). We use the UT and ET data to train the model and the PS data to validate its ability to generalize. 
The $\lambda$-PANN accurately captured the experimental data, with a loss on the order of $\sim 10^0$ and $\sim 10^{-1}$ for the Treloar and Heulliet datasets respectively (Figs. \ref{fig:rubber_experiment}b and \ref{fig:rubber_experiment}d). The number of active parameters in the final network is shown in blue. These are reduced due to L$^{0}$-regularization.

As a further indication of the accurate fit, we report $R^2$ values of the fits in Figs. \ref{fig:rubber_experiment}a and \ref{fig:rubber_experiment}c. In all cases $R^2>0.99$, highlighting that the trained constitutive model is also able to generalize well to unseen deformation modes (PS case).

\subsection{Finite element analysis}
\begin{figure}
    \centering
    \includegraphics[width=0.8\linewidth]{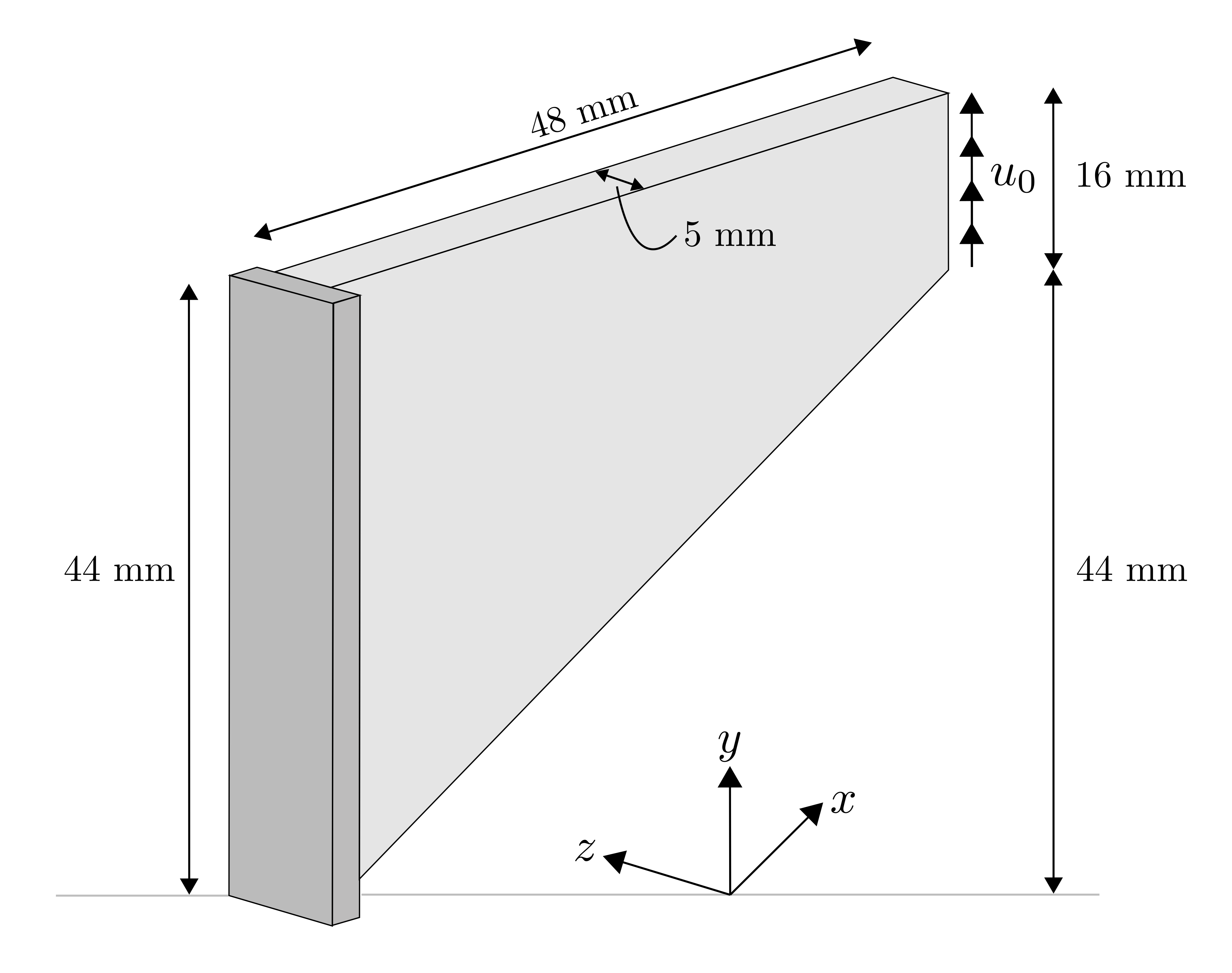}
    \caption{Structural Cook's membrane benchmark.}
    \label{fig:CooksMembrane}
\end{figure}

We implemented the $\lambda$-PANN architecture ($p=3$) trained on the first parameter set of the Ogden model (see Table \ref{tb:paramGeneralOgden}) into the open source finite element package \textit{Florence} \cite{florence}. Fig. \ref{fig:CooksMembrane} shows the benchmark boundary value problem known as Cook's membrane, which we chose to test our implementation of the $\lambda$-PANN model. 

\begin{figure}
    \centering
    \includegraphics[width=0.9\linewidth]{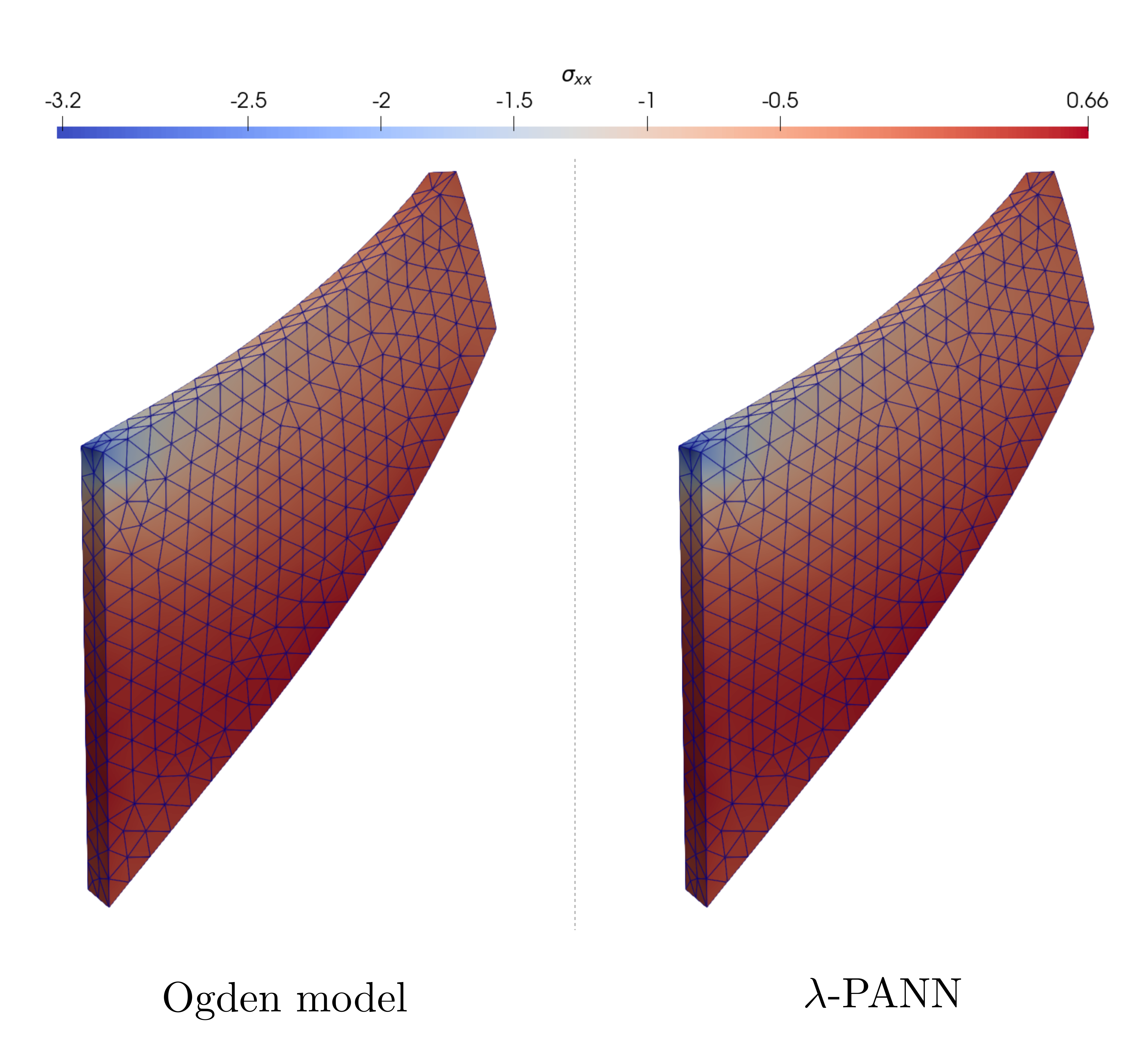}
    \caption{$\sigma_{xx}$ in the Cook's membrane model for the ground truth Ogden model (left) and the trained $\lambda$-PANN (right).}
    \label{fig:sigmaComparisonFE}
\end{figure}

\begin{figure}
    \centering
    \includegraphics[width=0.65\linewidth]{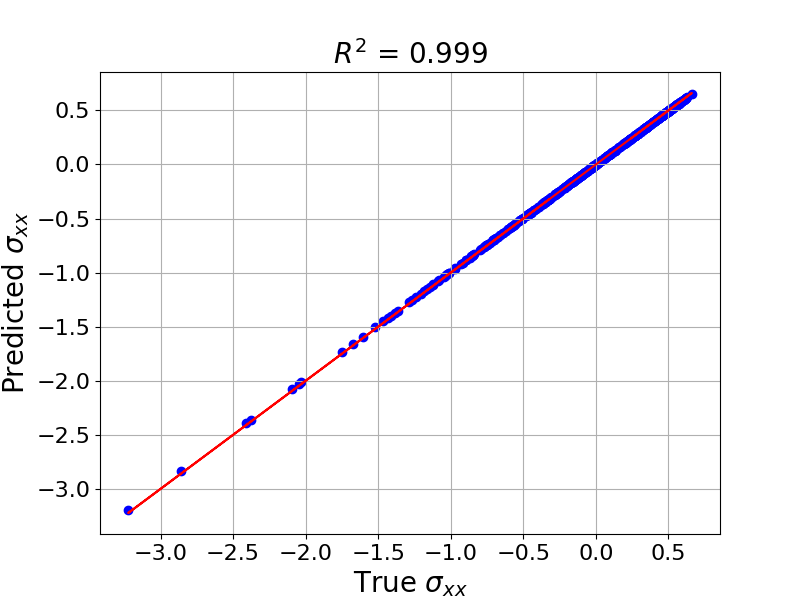}
    \caption{$R^{2}$ between true $\sigma_{xx}$ coming from the ground truth Ogden model and the predicted $\sigma_{xx}$ obtained by using the trained $\lambda$-PANN over all nodes of the Cook's membrane problem.}
    \label{fig:R2plotFE}
\end{figure}

Fig. \ref{fig:sigmaComparisonFE} shows side by side the stress contour prediction for the Cook's membrane problem using either the closed-form Ogden model Eq. (\ref{eq:genOgden}) or the $\lambda$-PANN architecture. Fig. \ref{fig:R2plotFE} further shows an accuracy plot between the stresses $\sigma_{xx}$ obtained from either the closed-form Ogden model or the $\lambda$-PANN model trained on the Ogden data. With $R^2=0.999$ it is clear that the implementation into the finite element package is correct. 

\section{Discussion}

In this manuscript, we present a physics-augmented data-driven constitutive model framework for hyperelastic materials in terms of principal stretches. By working with the eigenvalues of the right stretch tensor $\mathbf{U}$, the framework can satisfy objectivity. To satisfy polyconvexity, a sufficient condition for the existence of minimizers of the strain energy in boundary value problems (pending some growth conditions on the strain energy), the framework relies on the nesting of input-convex functions of $\mathbf{U}$, its cofactor $\text{cof}\mathbf{U}$, and its determinant $\det \mathbf{U}$. To guarantee isotropy and objectivity, we rely on permutation-invariant neural networks known as Holder deep power sets. In particular, we embed input-convex neural networks into the deep-set formulation, yielding symmetric convex functions of the eigenvalues of  $\mathbf{U}$, and of its cofactor $\text{cof}\mathbf{U}$. The framework accurately captures synthetic and experimental data, both within and outside of the training range of deformations. The framework can be readily implemented into finite element packages such as \textit{Jax-FEM} \cite{xue2023jax} or \textit{Florence}, as we do here \cite{florence}.


In recent years, numerous data-driven constitutive models have been proposed, primarily formulated in terms of the invariants of the right Cauchy-Green deformation tensor, $\mathbf{C}$. Notable exceptions include the constitutive artificial neural networks by St. Pierre et al. \cite{pierre2023principal} and the automated model discovery framework EUCLID \cite{flaschel2023automated}. However, these approaches extend the Ogden model by expressing the strain energy as a linear combination of multiple Ogden terms \cite{ogden1972large,destrade2022ogden}, rather than providing a truly general material representation in terms of principal stretches. While such an Ogden-like expansion can describe a broad class of materials, it remains a specialized form and may fail in certain cases \cite{anssari2022modelling}. Furthermore, the commonly used expansions, Eq. (\ref{eq:Ogden_compressible}), completely disregard the dependence on $\text{cof}\mathbf{F}$. 

Interestingly, the general Ogden material \cite{le1986incompressible,ciarlet1988three}, includes both contributions from $\mathbf{C}$ and $\text{Adj}\mathbf{C}$ (closely related to $\text{cof}\mathbf{F}$),
\begin{equation}
\Psi(\bm{F}) = \sum_{i=1}^{m} a_{i} \text{tr} (\bm{C}^{a_{i}/2}) + \sum_{j=1}^{n} b_{j} \text{tr} \left( (\text{Adj} \bm{C})^{b_{j}/2} \right) + \Gamma (\det \bm{F}),
\label{eq:Ogden_ciarlet}
\end{equation}

\noindent 
with
\begin{equation}
    \begin{aligned}
        \text{tr} (\bm{C}^{a/2}) &= \lambda_{1}^{a} + \lambda_{2}^{a} + \lambda_{3}^{a}  \\
        \text{tr} \left( (\text{Adj} \bm{C})^{b/2} \right) &= (\lambda_{1} \lambda_{2})^{b} + (\lambda_{1} \lambda_{3})^{b} + (\lambda_{2} \lambda_{3})^{b}.
    \end{aligned}
\end{equation}

However, this general expression is almost never used by practitioners. Our additive expansion, explored in Fig. \ref{fig:ablation}, is closely related to Eq. (\ref{eq:Ogden_ciarlet}), but instead of powers of the eigenvalues of $\mathbf{U}$ and $\text{cof}\mathbf{U}$ we employ the more general ICNN functions $\psi_{\mathbf{F}}$,$\psi_{\text{cof}\mathbf{F}}$. The results in Fig. \ref{fig:ablation} show that this additive decomposition is still limiting because it is unable to capture other nonlinear coupling terms. A fully nested structure was necessary to obtain a general functional form with excellent performance across all cases (see Fig. \ref{fig:ablation}). 

Invariant-based data-driven models have proven effective in representing a wide range of materials. However, it remained unexplored whether similarly flexible formulations could be formulated in terms of principal stretches. By capturing arbitrary convex symmetric functions with the convolution of deep Holder sets and ICNNs, it is unsurprising that the $\lambda$-PANN model performs comparably to the invariant-based $\mathcal{I}$-PANN when tested against synthetic and experimental data. Notably, $\lambda$-PANNs matched the performance of $\mathcal{I}$-PANN 
 even when trained on data generated from an invariant-based model. However, when the data originated from an Ogden model, $\lambda$-PANNs outperformed the  $\mathcal{I}$-PANN, particularly in extrapolation tasks. The $\lambda$-PANN architecture also captured the experimental data with high accuracy and generalized well to unseen loading modes. This suggests that while both frameworks are highly flexible and capable of capturing complex material behavior, $\lambda$-PANNs may offer a slight advantage

The performance of the $\lambda$-PANN was sensitive to the choice of hyperparameters. As mentioned above, the additive structure was unable to capture the synthetic examples, and a fully nested architecture was needed. For the fully nested $\lambda$-PANN, performance was influenced by the deep power set exponent $p = 1,2,3$ and whether the $p$-norm operated on the convex function $\phi$. Our results indicate that $p = 3$ with $\phi$ led to the best performance, achieving faster convergence compared to cases without $\phi$ and lower $p$. This raises the question of whether even larger values of $p$ could further improve performance.  As $p$ increases, the formulation approaches the $max$ norm, which is somewhat counterintuitive as it suggests that emphasizing the largest eigenvalue could yield the best constitutive model, but this remains to be tested.

While a fully nonlinear architecture was required, this did not imply an excessively large number of parameters. The intrinsic dimensionality of the data is relatively low. This is clear in synthetic data cases where the ground truth is represented with a small number of parameters, but it is expected in the experimental data cases as well \cite{fuhg2024extreme,linka2021constitutive}. In the synthetic cases, Fig. \ref{fig:NN_architecture} shows that even using a single layer in each of $\psi_{\mathbf{F}},\psi_{\text{cof}\mathbf{F}},\psi^{NN}$ achieves a low loss. This is likely related to the nonlinear nesting of functions in the architecture. In the experimental case, for which we used the architecture with two layers in each neural network, we applied $L_0$ regularization to enforce sparsity \cite{fuhg2024extreme}, reducing the number of parameters to approximately $10^1$. Thus, while fully nonlinear nesting is essential, achieving strong performance with a relatively small number of parameters is still possible.

We implemented the $\lambda$-PANN model into a finite element package. However, caution is required when computing the tangent stiffness, as it involves derivatives of the eigenvectors, which may not be uniquely defined when eigenvalues are repeated. This challenge is inherent not only to data-driven models but also to analytical models in terms of the principal stretches. Nevertheless, our finite element implementation showed no issues due to the usage of automatic differentiation on the strain energy function evaluated on numerically perturbed eigenvalues. 

This work is not without limitations. One limitation is the restriction to isotropic functions, whereas many soft materials, including biological tissues, exhibit anisotropy. Invariant-based models offer a natural advantage in this regard, as pseudo-invariants that account for material symmetries can be readily identified. Moving forward, an important direction would be to establish a corresponding set of \textit{invariants} in terms of eigenvalues and structural tensors. Even within the isotropic setting, however, the present formulation represents the most general principal-stretch-based model, going even beyond the general Ogden expansion Eq. (\ref{eq:Ogden_ciarlet}), which has remained the most common principal-stretch model for over fifty years. We anticipate that this work will stimulate further exploration of eigenvalue-based constitutive models, expand the use of data-driven strain energy functions, and lead to extensions accounting for dissipative mechanisms such as damage and viscoelasticity.

\section{Conclusion}
In this study, we introduced a physics-augmented, data-driven constitutive model framework for hyperelastic materials formulated in terms of principal stretches. By leveraging input-convex and permutation-invariant neural networks, we ensured polyconvexity, objectivity, and isotropy in our proposed $\lambda$-PANN architecture. Our model demonstrated strong performance across synthetic and experimental datasets, accurately capturing stress-strain relationships while generalizing well to unseen deformation modes. Notably, $\lambda$-PANNs outperformed invariant-based neural network models in extrapolation tasks, particularly when trained on Ogden-generated data.

Our finite element implementation further validated the efficacy of the $\lambda$-PANN model, yielding stress predictions nearly identical to those of the ground truth Ogden model in the Cook’s membrane benchmark. This highlights the practical applicability of the proposed architecture in real-world computational mechanics problems.


Ultimately, this study advances the use of polyconvex, data-driven approaches in computational mechanics, providing a flexible and physically consistent alternative to existing invariant-based formulations. We anticipate that this work will inspire further research into eigenvalue-based constitutive models, enhancing our ability to model complex material behavior with data-driven methods.

\section*{Data and Code}
After the acceptance of this manuscript the code and data will be made available under \url{https://github.com/FuhgJan/polyConvexPrincipalStrain}.

\bibliography{bib.bib}
\bibliographystyle{unsrt}
\end{document}